\documentclass[twocolumn,usenatbib]{mnras} 

\usepackage{subcaption}
\usepackage{graphicx}
\usepackage{float}
\usepackage{amsmath}
\usepackage{amssymb}	
\usepackage{siunitx}
\usepackage{physics}
\usepackage{xfrac}
\usepackage{comment}
\usepackage{hyperref}
\usepackage{makecell}
\usepackage{natbib}
\usepackage{dcolumn}
\usepackage{bm}
\usepackage[T1]{fontenc}
\usepackage{xcolor}     
\usepackage{enumitem}
\usepackage{eso-pic}

\setitemize{labelindent=0.5em,labelsep=0.1cm,leftmargin=0.1cm}

\def\pmb#1{\setbox0=\hbox{#1}%
    \kern-.025em\copy0\kern-\wd0
    \kern.05em\copy0\kern-\wd0
    \kern-.025em\raise.0433em\box0}

\def\ltsima{$\; \buildrel < \over \sim \;$}
\def\gtsima{$\; \buildrel > \over \sim \;$}
\def\simlt{\lower.5ex\hbox{\ltsima}}
\def\simgt{\lower.5ex\hbox{\gtsima}}
\def\p2Y{\;_2Y}
\def\m2Y{\;_{-2}Y}

\def\mk2{\mu {\rm K}^2}

\def\pmb#1{\setbox0=\hbox{#1}%
     \kern-.025em\copy0\kern-\wd0
     \kern.05em\copy0\kern-\wd0
     \kern-.025em\raise.0433em\box0}

\newcommand{\DG}[1]{\textcolor{orange}{#1}}

\definecolor{purple}{RGB}{156,81,182}

\newcommand{\rev}[1]{#1}

\begin{document}
\title[DES Y3: Blue shear]{Dark Energy Survey Year 3: Blue shear}
\AddToShipoutPictureBG*{%
  \AtPageUpperLeft{%
    \hspace{0.75\paperwidth}%
    \raisebox{-4.5\baselineskip}{%
      \makebox[0pt][l]{\textnormal{DES-2024-0855}}
}}}%

\AddToShipoutPictureBG*{%
  \AtPageUpperLeft{%
    \hspace{0.75\paperwidth}%
    \raisebox{-5.5\baselineskip}{%
      \makebox[0pt][l]{\textnormal{FERMILAB-PUB-24-0777-PPD}}
}}}%

\author[DES Collaboration]{
\parbox{\textwidth}{
\Large
J.~McCullough,$^{1,2,3}$\thanks{email: jmccullough@princeton.edu}
A.~Amon,$^{1}$
E.~Legnani,$^{4}$
D.~Gruen,$^{5,6}$
A.~Roodman,$^{2,3}$
O.~Friedrich,$^{7}$
N.~MacCrann,$^{8}$
M.~R.~Becker,$^{9}$
J.~Myles,$^{1}$
S.~Dodelson,$^{10,11}$
% ...
\begin{center} (DES Collaboration) \end{center}
}
}
\date{Accepted XXX. Received YYY; in original form ZZZ} 
\pubyear{\the\year{}}
\maketitle
\begin{abstract}
\noindent Modeling the intrinsic alignment (IA) of galaxies poses a challenge to weak lensing analyses. The Dark Energy Survey is expected to be less impacted by IA when limited to blue, star-forming galaxies. The cosmological parameter constraints from this \textsc{blue} cosmic shear sample are stable to IA model choice, unlike passive galaxies in the full DES Y3 sample, the goodness-of-fit is improved and the $\Omega_{\rm m}$ and $S_8$ better agree with the cosmic microwave background. Mitigating IA with sample selection, instead of flexible model choices, can reduce uncertainty in $S_8$ by a factor of 1.5.
\end{abstract}

\section{Introduction}\label{sec:intro}

Cosmic shear probes the growth of structures and the expansion of the Universe, providing a test of the standard cosmological model, $\Lambda$CDM. It measures the tiny distortions of the apparent shapes of background galaxies due to the weak gravitational lensing by foreground large- scale structure. Over the last two decades, this technique has matured and produced high-precision measurements of the amplitude of the \rev{matter} fluctuation spectrum, $S_8= \sigma_8 \sqrt{\Omega_{\rm m}/0.3}$\footnote{\rev{Here $\sigma_8$ is the root mean square linear amplitude of the matter fluctuation spectrum in spheres of radius $8 h^{-1} {\rm Mpc}$ $h$ is the universal expansion rate in km s$^{-1}$ Mpc$^{-1}$ scaled by a factor of 1/100 and $\Omega_{\rm m}$ is the present day matter density.}}  \citep{asgari_2021,Amon_2022, Secco_2022, dalal2023hyper, li2023hyper}. These constraints all show a preference for $S_8$ to be lower than that expected according to the best fit $\Lambda$CDM cosmology derived from the cosmic microwave background \textit{Planck} \citep{planck_2018_cosmo} (for a discussion see \citealt{amon_2022_s8}). 

Recent weak lensing surveys have demonstrated that \rev{their cosmological precision is limited by astrophysical systematic uncertainties \citep{Amon_2022, des+kids}}. Specifically, the dominant effects for cosmic shear are the modelling of the intrinsic alignment of galaxies (see  \citealt{troxel_rev, lamman_rev} for a review) and baryon feedback (see, for example \citealt{Bigwood}).
The Vera C. Rubin Observatory’s Legacy Survey of Space and Time\footnote{https://www.lsst.org} (LSST), ESA’s Euclid mission\footnote{https://www.euclid- ec.org}, and Roman Space Telescope\footnote{https://roman.gsfc.nasa.gov}, \citep{EuclidForecast, LSSTScience, Roman} are expected to lead to dramatic improvements in the statistical power of cosmic shear measurements. It is therefore critical to study these systematics to reduce their uncertainty and ensure that they do not bias cosmological inference \citep{lsstreqs2021}. 

Modern cosmic shear analyses commonly analyse their data using the Non-Linear Alignment model \citep[NLA; e.g.,][]{nla_desc,NLA_firstuse} and the Tidal Alignment and Tidal Torquing model \citep[TATT;][]{TATT}. NLA describes the linear tidal alignment of galaxies with the density field \citep{nla_desc}, including an \textit{ad hoc} non-linear correction to the linear matter power spectrum \citep{NLA_firstuse}, and a redshift dependence (`NLA-$z$', \rev{\citealt{nla_powerlaw}}). TATT extends NLA with the inclusion of a tidal torquing alignment mechanism. Critically, beyond a cost in precision, this IA model choice impacts the value of the reported cosmological constraints by $\sim$0.5$\sigma$ \citep{Secco_2022, Amon_2022, des+kids, Samuroff_2024} because uncertain IA model parameters are degenerate with cosmological parameters. 
As progress is made to build flexible feedback models that allow for cosmic shear analyses to utilise the full angular scale extent of measurements, IA modelling, which is more uncertain at small scales, comes into focus.
This highlights the importance to distinguish between IA models, or to devise new strategies to mitigate their impact.

The NLA and TATT models provide good fits to direct measurements of IA, which are constructed from weak lensing data that overlaps with accurate galaxy distances, such as spectroscopy \citep[e.g.,][]{eboss}. However, these measurements are limited to larger scales than cosmic shear analyses probe (\rev{6 Mpc$/h$ and $k$$\sim$1 for NLA and 2 Mpc$/h$ and $k$$\sim$3  for TATT}). We note that even though TATT provides accuracy on smaller scales, it has been found that cosmic shear data has a mild preference for the simpler NLA model over TATT \citep{Secco_2022}, though NLA breaks down from observation more quickly at smaller scales \citep{lamman2024}. To make progress on our understanding of IA, there has been effort to develop more sophisticated models based on the halo model \citep{Fortuna_2020} or other analytic approaches \citep[e.g.,][]{Vlah2020, Bakx2023, ChenKokron, Maion2024, mice, harnois_2022}, however many of these incur additional model parameters and exacerbate the precision cost \citep{Chen_2024}. Another avenue has been to test models in hydro-dynamical simulations, but these show a range of IA scenarios \rev{\citep{Delgado_2023, hydrosim_ia}}. 

It has been established from direct IA measurements that redder, elliptical galaxies at low redshift exhibit strong intrinsic alignment that is well described by NLA on large ($\gtrsim10\ h^{-1}$Mpc) scales \citep{hirata_2007,blazek_2011}, and that the strength of this alignment depends on galaxy luminosity \citep{Fortuna_measurements, eboss}. Bluer, star-forming galaxies have not been found to have any intrinsic alignment -- though direct measurements are limited \rev{\citep{eboss,wigglez_des,Johnston_2019}}. Indeed, a limitation of the NLA model is that it treats spiral galaxies and ellipticals equally, despite the very different physical mechanisms by which tidal forces may affect their shapes \citep{Catelan_2001,Mackey_2002}. Previously, cosmic shear data has been re-analysed with separate IA model parameters for star-forming (blue) and elliptical (red) galaxies in order to study the color-dependence of IA \citep{Samuroff_2019}. This analysis builds on this work but is driven by IA mitigation in order to realise cosmological posteriors that are stable to the IA model choice.   


In this paper, to bridge the limitations of IA model choice, model inaccuracies on small scales and precision-loss due to uninformed model space, we test a new approach to account for intrinsic alignments through sample selection. We construct a high-purity sample of star-forming galaxies from the Dark Energy Survey (DES) Year 3 (Y3) lensing data, that \textit{a priori} is expected to have no or minimal IA consistent with presently uncertain direct measurements, therefore circumventing the need for an IA model in the cosmic shear analysis \citep{Krause_2015}. We calibrate this data, measure the 2-point shear correlation functions and analyse them to demonstrate a cosmic shear cosmology result that is stable to the choice of IA model. Without the additional concern of an accurate model on small angular scales, we analyse the full measurements using a flexible model for baryon feedback.  We examine the remaining mixed sample with a high fraction of passive, redder elliptical galaxies to test the IA model suitability.  In Sec.~\ref{sec:method} we present the methodology for selecting an `IA-clean' sample and summarize the subsequent shear measurement. In Sec.~\ref{sec:model}, we outline the model choices and framework. The main results of this paper are contained in Sec.~\ref{sec:results} and we discuss the implications of our findings in Sec.\ref{sec:discussion}.

\section{Methodology}
\label{sec:method}
\begin{figure*} 
    \caption{The calibrated redshift distributions for the \textsc{blue}, \textsc{red}, and \textsc{full} samples (black) for each DES Y3 redshift bin, where the SOM selection is shown in the insets (cells not selected for the redshift bin are filled in black).}
    \centerline{\includegraphics[width=1.0\textwidth]{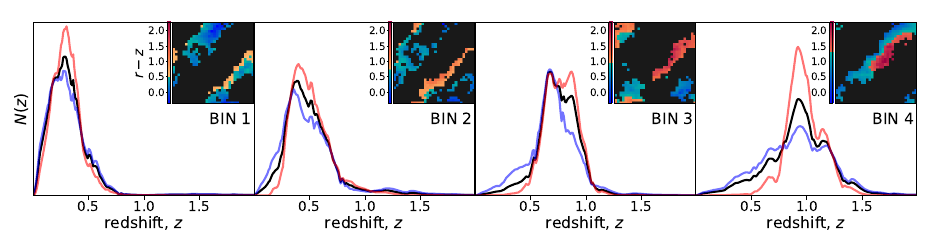}}
    \label{fig:fidcolor}
\end{figure*}

This work expands upon the DES Y3 cosmic shear analysis \citep{Amon_2022, Secco_2022}. DES Y3 weak lensing data spans 4143 deg$^2$ with $riz$ photometry and \textsc{metacalibration} shapes  \citep{Gatti_2021} and is divided into four redshift bins \citep{Myles_2021}. For each redshift bin, we select a high-purity star-forming sample, \textsc{blue}, described in Sec.~\ref{sec:color}, leaving a predominantly elliptical sample, \textsc{red}. We calibrate the redshift distributions of the data (Fig.~\ref{fig:fidcolor}) following the \textsc{sompz} methodology, which hinges on deep fields with overlapping near-infrared observations \citep{Hartley_2021} and survey characterization with \textsc{balrog} \citep{Everett_2022}, used in a Self Organizing Map (SOM) framework. We repeat the shear calibration for the \textsc{blue} and \textsc{red} samples \citep{MacCrann_2021}, which is based upon image simulations. This calibration and any customization of the DES Y3 strategy is described further in App.~\ref{sec:data}. The selection, properties and calibration parameters of the sample are quantified in Tab.~\ref{tab:data_properties}. Measurements of the \textsc{blue} and \textsc{red} shear 2-point functions are presented in Sec.~\ref{sec:cosmic_shear}.


\subsection{Selection of star-forming galaxies}
\label{sec:color}

\begin{table*}
    \caption{The statistics of the \textsc{blue} and \textsc{red} galaxy subsamples of the DES Y3 data for each redshift bin, including the effective number density $n_{\rm eff}$ (gal/arcmin$^2$), the \rev{average} per-component shape noise, \rev{$\langle\sigma_{e}\rangle$}, the calibrated mean redshift, \rev{$\langle z\rangle$}, the uncertainty in the mean redshift (\rev{adopted from \citealt{Myles_2021}}), $\Delta z$, and the shear calibration parameter, $m$. For each bin, the subsamples are divided by the $r-z$ selection value, tuned and characterized by the fraction of passive, red galaxies inferred from the optical and near-infrared deep field data using three techniques (see Sec.~\ref{sec:color}).}
    \label{tab:data_properties}
\begin{center}
\resizebox{\textwidth}{!}{\begin{tabular}{cccccccccccc}
\hline
\hline
Bin & \vtop{\hbox{\strut\centering{$r-z$}}\hbox{\strut \centering Selection}} & No. objects & \rev{\vtop{\hbox{\strut\centering{Frac. of}}\hbox{\strut \centering DES Y3}}} & $n_\text{eff}$ & \rev{$\langle\sigma_{e}\rangle$} & \rev{$\langle z\rangle $} & \vtop{\hbox{\strut\centering\% Red}\hbox{\strut \centering(\textsc{bagpipes})}}  & \vtop{\hbox{\strut\centering\% Red}\hbox{\strut\centering (\textsc{ea}z\textsc{y})}} & \vtop{\hbox{\strut\centering\% Red}\hbox{\strut\centering(balmer)}} & $\Delta z$ & $m\times100$ \tabularnewline
\hline 
\textsc{blue}, 1 & $< 0.50$ & $18\,031\,829$ & 0.725 & 1.054 & 0.255 & 0.3556 & 1.14 & 1.51 & 4.77 & 0.018& $-1.29 \pm 0.91$ \tabularnewline 
\textsc{blue}, 2 & $< 0.75$ & $16\,670\,470$ & 0.660 & 0.988 & 0.282 & 0.5175 & 2.71 & 3.13 & 9.21 & 0.015 & $-1.80 \pm 0.78$ 
\tabularnewline
\textsc{blue}, 3 & $< 0.95$ & $12\,233\,530$ & 0.493 & 0.726 & 0.275 & 0.6994 & 2.81 & 2.94 & 7.79 & 0.011 & $-2.03 \pm 0.76$
\tabularnewline
\textsc{blue}, 4 & $< 1.30$ & $18\,130\,765$ & 0.718 & 1.073 & 0.325 & 0.8994 & 2.22 & 2.88 & 6.51 & 0.017 & $-3.56 \pm 0.76$
\tabularnewline
\hline 
\textsc{red}, 1 & $> 0.50$ &$6\,850\,889$ & 0.275 & 0.439 & 0.227 & 0.3025 & 18.14 & 19.20 & 36.20 & 0.018 & $-0.89 \pm 0.91$
\tabularnewline
\textsc{red}, 2 & $> 0.75$ &$8\,554\,173$& 0.340 & 0.548 & 0.246 & 0.5116 & 27.92 & 26.98 & 55.05 & 0.015& $-1.78 \pm 0.78$
\tabularnewline
\textsc{red}, 3 & $> 0.95$ & $12\,597\,837$ & 0.507 & 0.780 & 0.242 & 0.7757 & 25.63 & 21.77 & 46.75 & 0.011 & $-2.37 \pm 0.76$
\tabularnewline
\textsc{red}, 4 & $> 1.30$ & $7\,134\,533$ & 0.282 & 0.418 & 0.294 & 0.9851 & 30.46 & 29.25 & 42.12 & 0.017 & $-3.15 \pm 0.76$
\tabularnewline
\hline
\end{tabular}}
\end{center}
\end{table*}

\par
For each of the four DES Y3 redshift bins \citep{Myles_2021}, we select on $r-z$ color of a DES Y3 SOM cell, visualized in the inset panels in Fig.~\ref{fig:fidcolor}. The selection is designed to maximize the purity of the \textsc{blue} sample and therefore minimize intrinsic alignment.  While for the vast majority of galaxies we have only \textit{riz} colors alongside their shapes, the deep-fields contain a wealth of information in \textit{ugrizJH}$K_s$ photometry that can aid and characterize our fiducial selection. The selection is tuned to produce sufficient statistics for cosmic shear, and red galaxies are excluded with high fidelity using three cross-checks informed by deep field inference of passive galaxy fractions: (1) SED fitting with \textsc{bagpipes} \citep{Carnall_2018}, \rev{which models the emission from galaxies and allows us to select red galaxies with negligible star formation rates}, (2) SED fitting with \textsc{eazy} \citep{eazy_2008}, \rev{which allows for selection on galaxies dominated by passive galaxy templates --complementary to \textsc{bagpipes} as the selection is on template coefficients and not derived galaxy properties}, and (3) \rev{observed color in the filters bracketing the Balmer break at the inferred photometric redshift} (for more detail, see App.~\ref{app:color}).
The $r-z$ boundary that defines the \textsc{blue} sample, \rev{the fraction of the sample per bin} and the purity estimated by each method is given in Tab.~\ref{tab:data_properties}.
These indicate that the \textsc{blue} sample is $\lesssim$ 3\% contaminated with red galaxies (as the Balmer break method overestimates the total red fraction), \rev{and comprises 65 percent of the Y3 sample}. 
\rev{We note that even the \textsc{red} sample is estimated to be only $\sim$20-30\% red.}

\rev{Compared to previous work, our emphasis is on purity in the \textsc{blue} sample to derive an `IA-clean sample', rather than have a well-defined red and blue split. Our sample selection is improved and better controlled, aided by the 8-band deep field information and the SOMPZ framework, without a reliance on BPZ templates compared to optical-only information.}



\subsection{Cosmic shear modeling}\label{sec:model}
\begin{figure*}
    \centering
    \includegraphics[width=0.98\textwidth]{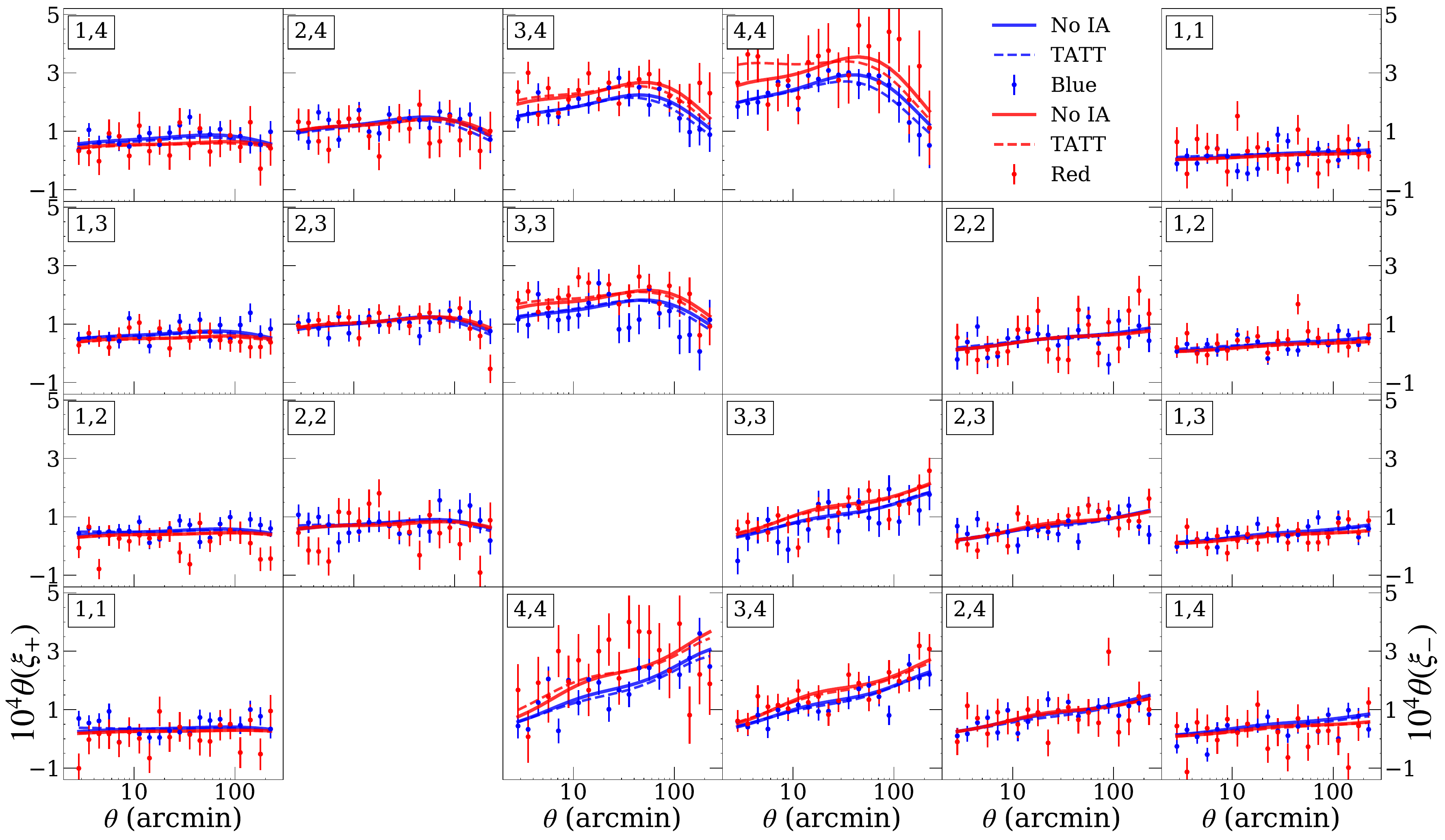}
    \caption{Cosmic shear two-point correlation measurements for each redshift bin pair, $\xi_{\pm}$, for the \textsc{blue} and \textsc{red} samples. 
    The error bars represent the square root of the diagonal of the analytic covariance matrix. The solid lines represent the best model fit for $\Lambda$CDM with no intrinsic alignment, which is preferred by both selections, and the dashed lines represent the best fit for TATT, the most complex IA model choice considered. Notable shifts are at small scales in the autocorrelation for bin 4 -- marked [4,4]. For residuals, see Fig.~\ref{fig:cosmicshear_resid}.}
    \label{fig:cosmicshear}
\end{figure*}

The cosmic shear measurements, $\xi_{\pm}$, can be written as a function of a convergence $\kappa$ power spectrum, which results from the 3D non-linear matter power spectrum computed at a given choice of cosmological parameters. For the modelling and analysis, we closely follow the DES Y3 choices \citep{Amon_2022, Secco_2022}, \rev{detailed in App.~\ref{app:cosmology} with salient updates summarised as}: 
\begin{itemize}
\setlength\itemsep{-1em}
\item Following \cite{des+kids}, we model the non-linear power spectrum using \textsc{HMCode} \citep{mead_2020} and we account for baryon feedback by marginalizing over the log$_{10}$ $T_{\rm AGN}$ parameter. Following the model tests in \cite{Bigwood}, we choose a flexible prior of 7.6-8.3 to account for extreme feedback scenarios. \\
\item We analyse the full range of angular scales plotted in  Fig.~\ref{fig:cosmicshear}. The DES Y3 scale cuts (defined in \citealt{Amon_2022, Secco_2022}) are primarily designed to mitigate the impact of baryon feedback,  however, even with conservative feedback modeling, the fidelity of the NLA and TATT intrinsic alignment models on small-scales is uncertain. A key advantage of mitigating IA through sample selection with a \textsc{blue} sample demonstrated in this paper is the ability to safely exploit the small-scale cosmic shear measurements. \\
\item We assume three neutrino species with two massless states and one massive state and a mass of $0.06 {\rm eV}$, which has been shown to have a negligible impact \citep{Secco_2022, des+kids, Bigwood}.
\end{itemize}

Here we briefly describe the intrinsic alignments models. The observed ellipticity of a galaxy, $\epsilon_{\rm obs}$, can be approximately written as having linear contributions from cosmic shear, $\gamma_{\mathrm{lens}}$, the intrinsic alignment, $\gamma_{\mathrm{IA}}$ and its true shape, $\epsilon_{\mathrm{no IA}}$: $\epsilon_{\mathrm{obs}}$ \rev{$\approx$} $\epsilon_{\mathrm{no IA}} + \gamma_{\mathrm{lens}} +\gamma_{\mathrm{IA}}$. The orientation of $\epsilon_{\mathrm{no IA}}$ is uncorrelated between any two galaxies, but $\gamma_{\mathrm{lens}}$ and $\gamma_{\mathrm{IA}}$ are spatially coherent, and correlated with one another, for pairs of galaxies. App.~\ref{app:cosmology} details how this propagates into our measurements as a function of galaxy pair separation.


The TATT IA model \citep{TATT}, which was the fiducial choice for the DES Y3  \citep{Amon_2022,Secco_2022} \rev{and HSC Y3 \citep{dalal2023hyper,li2023hyper} analyses,} allows for more complexity than the NLA, favored in \citep[e.g.,][]{asgari_2021, des+kids}. Within TATT, the responses to large-scale tidal fields are encapsulated in three parameters\rev{:} $A_1$, $A_2$, and $A_{1\delta}$. These correspond to a linear response of galaxy shape to the tidal field (tidal alignment), a quadratic response (tidal torquing), and a response to the product of the density and tidal fields (density weighting):
\begin{equation}
\gamma_{ij}^{IA} = A_{1\delta}(\delta\times s_{ij}) +A_1 s_{ij}
+A_2 \left( \sum_{k=0}^2s_{ik}s_{kj} - \frac{1}{3}\delta_{ij}s^2 \right),
\end{equation}
where $\delta$ and $s$ describe the density and tidal fields, respectively.
The redshift evolution of $A_1$ and $A_2$ is parameterized as a power law, governed by $\eta_1$ and $\eta_2$, and given by,
\begin{align}
  {}
  & A_{1\delta}(z) = b_{\rm ta} A_1(z)\, ,\\
  &  A_1(z) = -a_1 \bar{C}_1 \frac{\rho_{\rm m} }{D(z)} \left( \frac{1+z}{1+z_p}\right )^{\eta_1}\, , \\
   &  A_2(z) =   5 a_2 \bar{C}_1 \frac{\rho_{\rm m} }{D(z)^2} \left( \frac{1+z}{1+z_p}\right )^{\eta_2}\, , 
\end{align}
\noindent
where $D(z)$ is the linear growth factor, $\rho_{\rm m} = \Omega_{\rm m} \rho_{\rm crit}$ is the matter density, $\bar{C}_1$ is a normalisation constant by convention fixed at a value of $5\times10^{-14} {\rm M_{\odot}}h^{-1}{\rm Mpc}^2$ \citep{brown_2002}, and $z_p$ is a pivot redshift, which we fix to the value $0.62$ \citep{Troxel_2018}. In the absence of informative priors, the analysis marginalizes over all five IA parameters that govern the amplitude and redshift dependence of the signal, $(a_1, a_2, \eta_1, \eta_2, b_{\rm ta})$, with non-informative flat priors summarised in App.~\ref{app:cosmology}. Other IA models are nested within the TATT framework, where the simplest, single parameter model corresponds to only $a_1$ being varied \rev{(NLA)}. Uncertainty or mismodeling of IA can also enter our analysis as shifts of the mean of the redshift distributions, discussed further in  App.~\ref{app:ia_pz}.

\subsection{Cosmic shear measurements}\label{sec:cosmic_shear}

\begin{figure}
    \centering
    \includegraphics[width=0.9\columnwidth]{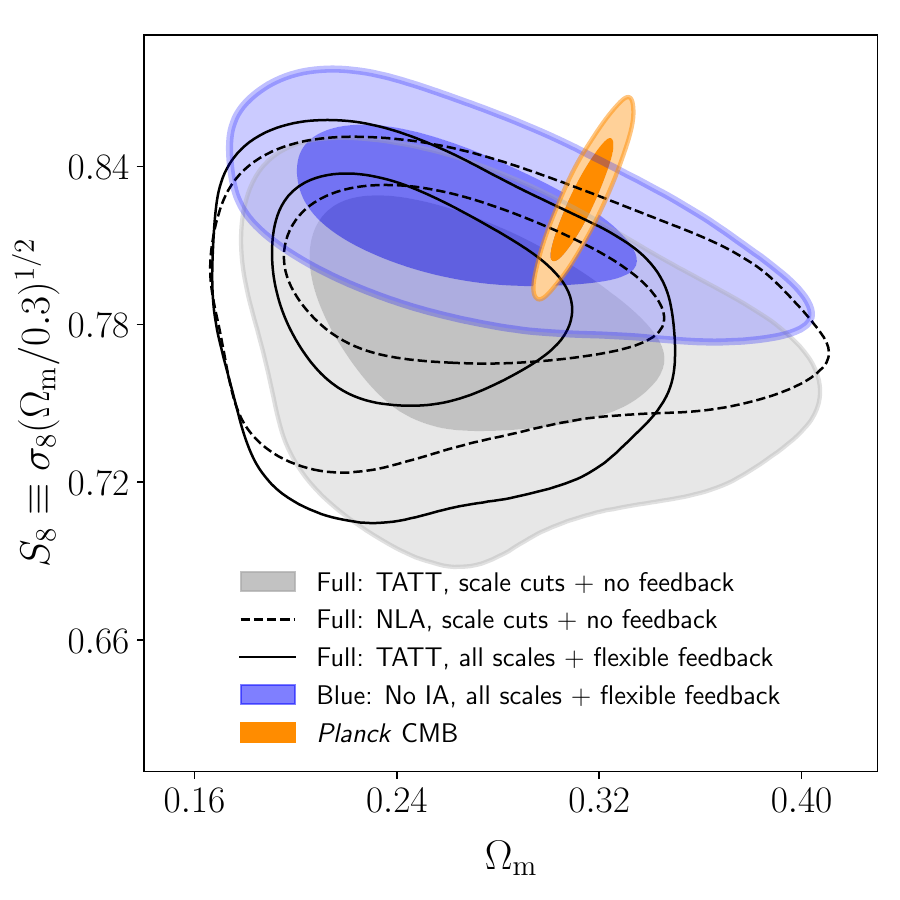}
    \caption{Marginalized posteriors for $\Omega_{\rm m}$ and $S_8$ derived from cosmic shear measurements using only \textsc{blue} galaxies, designed to mitigate the effects of IA (blue). These are analyzed with a flexible baryon feedback model, and compared to the \textsc{full} DES Y3 sample using the same feedback model and the TATT IA model (black line), producing a lower value for $S_8$ and $\Omega_m$. We also compare to the Y3 approach to exclude small-scale measurements of \textsc{full} sample and model IA using TATT (gray filled) and NLA (black dashed). \rev{At a cost of removing 35\% galaxies}, the \textsc{blue} shear $S_8$ constraint is \rev{1.5$\times$} improved precision and shifts \rev{1.5$\sigma$} higher. 
    The \textit{Planck} TTTEEE constraints are shown in orange \citep{EfstathiouGratton:2021}.}
    \label{fig:scale_cut_comparisons}
\end{figure}

We follow the DES Y3 methodology \citep{Amon_2022,Secco_2022,3x2pt} to compute the cosmic shear measurements and covariance for the \textsc{blue} and \textsc{red} samples. The shear two-point correlation function is determined by averaging over galaxy pairs, $(1,2)$, separated by an angle, $\theta$, for two redshift bins, $(i,j)$, as
\begin{equation}
\xi_{\pm}^{ij}(\theta) = \frac{\sum_{1,2} w_1 w_2 [\epsilon_t^i \epsilon_t^j \pm \epsilon_{\times}^i \epsilon_{\times}^j]}{\sum_{1,2} R_a R_b w_1 w_2 },
\end{equation}
in terms of their measured radial and tangential ellipticities, $\epsilon_{\times}$ and $\epsilon_t$, and where galaxy pairs are weighted by an inverse variance of the shear information they contain, $w$, and a metacalibration response, $R$, that corrects for biases due to sample selection and shear measurement \citep{Gatti_2021}. The measurements are shown in Fig.~\ref{fig:cosmicshear} alongside their best-fit models using TATT and no IA model.\footnote{In the Y3 fiducial analysis we found an outdated tomographic binning, and as a consequence of this analysis we provide a marginally updated full (un-split by color) datavector. These updates change the $\chi_{min}^2$ and cosmology results to a degree, and the \textsc{full} sample in this analysis is not analogous to the Y3 fiducial as a result.}

\begin{figure*}
    \centering
    \input{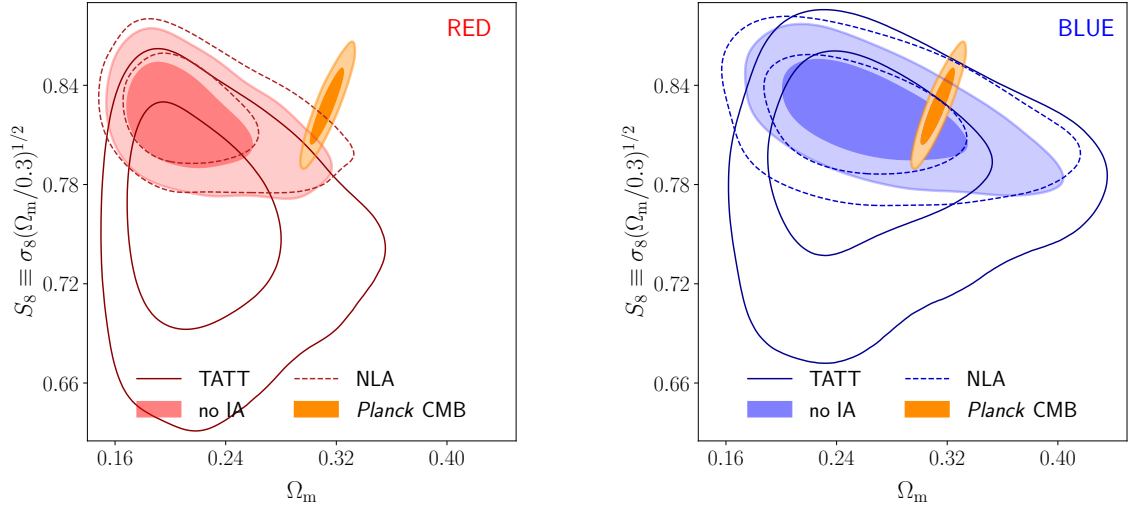}
    \caption{\label{fig:s8-om} Marginalized posteriors for $\Omega_{\rm m}$ and $S_8$ using 
    the impure \textsc{red} (left) and pure \textsc{blue} (right) samples. For each sample, we consider an analysis with no intrinsic alignment model (\rev{filled}), with the single parameter NLA model (\rev{dashed, unfilled}) and the TATT model (solid, unfilled). For reference, we show the \textit{Planck} TTTEEE likelihood \citep[orange,][]{EfstathiouGratton:2021}. The inner and outer contours correspond to 68\% and 95\% confidence levels, respectively.}
\end{figure*}

\newlength{\mytabcolsep}
\setlength{\tabcolsep}{6pt}

\begin{table*}
\resizebox{\textwidth}{!}{
\begin{tabular}{ccccccccccccccc}
\hline
\hline
Model & $S_{8, \mathrm{mean}}$ & $\Omega_{\mathrm{m}, \mathrm{mean}} $ & $S_{8, \mathrm{MAP}}$ & $\Omega_{\mathrm{m}, \mathrm{MAP}} $& \makecell{Evid.\\Ratio} & $\chi_{\mathrm{min}}^2$  & $\chi_{\mathrm{red}}^2$ & $\delta S_8$ & $\delta \Omega_\mathrm{m}$ & $\sqrt{\delta S_8^2 + \delta \Omega_\mathrm{m}^2}$\\
\hline
 \textsc{full} NLA ($a_1$) & $ 0.811^{+0.016}_{-0.019} $ & $0.255^{+0.031}_{-0.051} $ & 0.825 & 0.230 & {1.0} & {403.3} & {1.0} & {0.72} & {1.29} & {1.48} \\ 
  \textsc{full} NLA-$z$ ($a_1,\eta_1$) & $ 0.804^{+0.023}_{-0.018} $ & $0.249^{+0.029}_{-0.048} $ & 0.823 & 0.208 & {1.5} & {404.0} & {1.0} & {0.87} & {1.51} & {1.74} \\ 
  \textbf{(2) \textsc{full} TATT} & $ 0.788^{+0.033}_{-0.025} $ & $0.247^{+0.030}_{-0.044} $ & 0.778 & 0.240 & {0.8} & {399.5} & {1.0} & {1.16} & {1.72} & {2.08} \\ 
  \textsc{full} No IA & $ 0.811^{+0.017}_{-0.019} $ & $0.255^{+0.028}_{-0.042} $ & 0.826 & 0.228 & {20.4} & {403.9} & {1.0} & {0.72} & {1.53} & {1.69} \\ 
  \textsc{full} NLA-$z$, Y3 Scale Cuts\textsuperscript{\textdagger} & $ 0.793^{+0.027}_{-0.019} $ & $0.267^{+0.035}_{-0.059} $& 0.801 & 0.250 &  & {299.8} & {1.1} & {1.21} & {0.92} & {1.52} \\ 
  \textbf{(3) \textsc{full} TATT, Y3 Scale Cuts}\textsuperscript{\textdagger} & $ 0.777^{+0.035}_{-0.024} $ & $0.273^{+0.035}_{-0.052} $& 0.776 & 0.308 &  & {293.1} & {1.1} & {1.42} & {0.90} & {1.68} \\ 
 \hline 
  \textsc{red} NLA ($a_1$) & $ 0.822^{+0.022}_{-0.022} $ & $0.218^{+0.017}_{-0.045} $ & 0.868 & 0.161 & {1.0} & {426.4} & {1.1} & {0.24} & {2.40} & {2.41} \\ 
  \textsc{red} NLA-$z$ ($a_1,\eta_1$) & $ 0.820^{+0.021}_{-0.023} $ & $0.221^{+0.020}_{-0.048} $ & 0.857 & 0.162 & {0.7} & {424.5} & {1.1} & {0.28} & {2.24} & {2.25} \\ 
  \textsc{red} TATT & $ 0.753^{+0.050}_{-0.035} $ & $0.224^{+0.022}_{-0.049} $ & 0.789 & 0.201 & {3.7} & {419.2} & {1.1} & {1.57} & {2.16} & {2.67} \\ 
  \textsc{red} No IA & $ 0.818^{+0.019}_{-0.020} $ & $0.215^{+0.017}_{-0.041} $ & 0.862 & 0.167 & {5.9} & {425.1} & {1.1} & {0.39} & {2.76} & {2.79} \\ 
  \textsc{red} NLA-$z$, Y3 Scale Cuts\textsuperscript{\textdagger} & $ 0.792^{+0.029}_{-0.028} $ & $0.238^{+0.021}_{-0.061} $& 0.826 & 0.182 &  & {302.8} & {1.1} & {1.07} & {1.43} & {1.79} \\ 
  \textsc{red} TATT, Y3 Scale Cuts\textsuperscript{\textdagger} & $ 0.762^{+0.049}_{-0.029} $ & $0.231^{+0.023}_{-0.055} $& 0.752 & 0.175 &  & {301.5} & {1.1} & {1.41} & {1.73} & {2.23} \\ 
 \hline 
  \textsc{blue} NLA ($a_1$) & $ 0.820^{+0.023}_{-0.024} $ & $0.262^{+0.035}_{-0.062} $ & 0.819 & 0.254 & {1.0} & {380.2} & {1.0} & {0.28} & {0.95} & {0.99} \\ 
  \textsc{blue} NLA-$z$ ($a_1,\eta_1$) & $ 0.811^{+0.032}_{-0.024} $ & $0.260^{+0.038}_{-0.071} $ & 0.821 & 0.238 & {1.6} & {380.6} & {1.0} & {0.49} & {0.91} & {1.04} \\ 
  \textsc{blue} TATT & $ 0.793^{+0.044}_{-0.030} $ & $0.266^{+0.034}_{-0.064} $ & 0.813 & 0.269 & {0.6} & {375.4} & {1.0} & {0.79} & {0.87} & {1.18} \\ 
  \textbf{(1) \textsc{blue} No IA} & $ 0.822^{+0.019}_{-0.020} $ & $0.268^{+0.031}_{-0.056} $ & 0.810 & 0.260 & {12.7} & {380.9} & {1.0} & {0.23} & {0.95} & {0.98} \\ 
  \textsc{blue} NLA-$z$, Y3 Scale Cuts\textsuperscript{\textdagger} & $ 0.770^{+0.059}_{-0.025} $ & $0.274^{+0.045}_{-0.070} $& 0.696 & 0.181 &  & {259.0} & {1.0} & {1.10} & {0.64} & {1.27} \\ 
  \textsc{blue} TATT, Y3 Scale Cuts\textsuperscript{\textdagger} & $ 0.754^{+0.070}_{-0.034} $ & $0.292^{+0.044}_{-0.072} $& 0.771 & 0.351 &  & {252.9} & {0.9} & {1.22} & {0.35} & {1.26} \\ 

\hline
\end{tabular}
}
\caption{Cosmological results for the \textsc{full}, \textsc{red}, and \textsc{blue} samples as mean and \textit{maximum a posteriori} (MAP) values, with the goodness of fit, $\chi^2_{\rm red}=\chi^2_{\rm min}/(N_{\rm data}-N_{\rm par})$. $N_{\rm data}=400$ for the full scale extent of the measurements and 273 with scale cuts, and we estimate the effective number of free parameters as $N_{\rm par}\sim[5, 3, 2.5, 2]$ for [TATT, NLA-$z$, NLA and no IA], respectively, as per \protect\cite{eff_dof,Secco_2022}. The $\chi^2_{\rm min}$ are substantially improved for the \textsc{blue} sample (lower than that for \textsc{full}, \textsc{red}  by $\sim20,40$) and the \textsc{blue} cosmological constraints are more stable across IA model choices. The Bayesian evidence ratio with respect to NLA shows that no IA model is preferred for all samples. The differences compared to \textit{Planck} \protect\citep{EfstathiouGratton:2021} (Tab.~17, $S_8 = $0.828(16), $\Omega_{\mathrm{m}} = $0.3135(81) updated from \protect\citealt{planck_2018_cosmo}), $\delta S_8, \delta\Omega_{\rm m}$, are smallest for the \textsc{blue} sample.}
\label{tab:cosmologyresults}
\end{table*}
\setlength{\mytabcolsep}{\tabcolsep} 

\section{Results \& Discussion}\label{sec:results}

We determine the posterior of cosmological and nuisance parameters by sampling the prior and likelihood function. 
The 2D marginalized constraints on the matter fluctuation amplitude $S_8$, and matter density $\Omega_{\rm m}$, for the \textsc{blue} sample are summarised in Fig.~\ref{fig:scale_cut_comparisons}. The mean values of $S_8$ and their 68\% credible intervals for three IA model variants are found to be 
\begin{align}
\label{eqn:S8results}
\begin{aligned}
{\rm no \,\,IA:}\,\,\, &S_8 &=&\,\,\,  0.822^{+0.019}_{-0.020} \\  
{\rm NLA:}\,\,\, &S_8 &=&\,\,\, 0.820^{+0.023}_{-0.024} \\  
{\rm TATT:}\,\,\, &S_8 &=&\,\,\,  0.793^{+0.044}_{-0.030} \,,\\  
\end{aligned}
\end{align}
\noindent 
which are consistent within 0.5$\sigma$. Table \ref{tab:cosmologyresults} lists the constraints on $S_8$ and $\Omega_{\rm m}$ for all three samples. 
We compare our proposed `no IA' blue sample cosmic shear analysis to the results from the DES Y3 approach, which discarded small-scale measurements ($\Lambda$CDM-Optimized scale cuts) to limit the impact of astrophysical systematics, primarily baryon feedback, and used the TATT IA model. This has the added value to minimizing the impact of IA,  \citep[e.g.,][]{des+kids,Secco_2022, Amon_2022, Troxel_2018}. For reference, we compare to the CMB constraint from \textit{Planck} TTTEEE\footnote{\rev{The} high multipole likelihood attained from combining the temperature power spectra (TT), temperature-polarization E-mode cross spectra (TE) and polarization E-mode power spectra (EE).} $\Lambda$CDM analysis \citep[][]{EfstathiouGratton:2021}. Our approach to mitigate IA through a pure blue selection and a flexible baryon model allows for safe use of all angular scales of the measurement. It produces almost a  factor of two improvement on the $S_8$ constraint, despite having lower statistical weight -- due in part to the advantage of fewer model parameters. While the use of scale cuts \rev{on the \textsc{full} sample} does reduce the \rev{slight disagreement in} $\Omega_{\rm m}$ between the \textsc{full} and \textsc{blue} samples, the $S_8$ value falls below the \textsc{blue} result by 1.3$\sigma$.

In Fig.~\ref{fig:s8-om} we contrast the cosmological constraints derived from analyses of the  \textsc{blue} and \textsc{red} samples using three IA model variants: no IA model, NLA and TATT. The \textsc{blue} sample shows stability across IA model choices, whereas  the $S_8$ values obtained for the \textsc{red} in the three IA model cases can vary by up to 1.4$\sigma$. The \textsc{blue} shear analysis is also best described by the $\Lambda$CDM model, with the goodness-of-fit for each sample and model quantified in Table \ref{tab:cosmologyresults}. \rev{We approximately report $\chi_{\mathrm{red}}^2$ based upon the estimation of effective degrees of freedom in \cite{Secco_2022,Amon_2022}, from the minimum $\chi_2$ in the chain}. The \textsc{blue} sample has $p \geq 0.70$ across all model choices, while \textsc{full} ($p \leq 0.43$) and \textsc{red} ($p \leq 0.20$) are low in comparison, indicating their data are less well described by any model choice.  

We compare our constraints on $S_8$ and $\Omega_{\rm m}$ to the \textit{Planck} CMB \citep{EfstathiouGratton:2021} parameters and report shifts as $\delta S_8$ and $\delta\Omega_{\rm m}$\footnote{For simplicity, we quantify the shifts in the measured values of $S_8$ and $\Omega_{\rm m}$ between two analyses using $\delta X = \Delta X/[(\sigma_{X}^1)^2+(\sigma_{X}^2)^2]^{1/2}$, where $\Delta X$ is the difference between the respective mean values and $\sigma_{X}^1$,$\sigma_{X}^2$ are the standard deviations of the two constraints.}. We find that \textsc{blue} is in moderately improved statistical agreement with \textit{Planck} compared to \textsc{red} and \textsc{full}, in all IA model cases. In particular, \textsc{blue}: no-IA has $\delta S_8 = 0.23$, while the the full has $\delta S_8 = 0.72$ and the impure \textsc{red} sample has $\delta S_8 = 0.39$. These improvements are mild until we consider $\Omega_{\rm m}$. Cosmic shear analyses that include the small-scale measurements have reported low values of the matter density parameter compared to \textit{Planck} \citep[e.g.,][]{GarciaGarcia, Bigwood}, which is stable to baryon feedback models \citep{Bigwood}. The \textsc{blue} $\Omega_{\rm m}$ constraint is consistent with \textit{Planck} within 1$\sigma$. For \textsc{full} and \textsc{red}, where we expect galaxies to intrinsically align, we find $\delta \Omega_{\rm m} > 1.5\sigma,\ 2.1\sigma$, with no improvement when considering more complex IA models (see Table \ref{tab:cosmologyresults}). Given that IA produces a scale dependent effect, and that the IA models have not been sufficiently validated on the scales that cosmic shear is most sensitive to, it is plausible that IA mismodelling would result in a biased value of $\Omega_{\rm m}$, in addition to $S_8$.

For all samples, Bayesian evidence ratios prefer no intrinsic alignment. The strength of that model preference depends on sample, and is weakest for the \textsc{red} sample. In the pure \textsc{blue} sample, it is strongly preferred (Bayes Evidence ratio $K = 7.9\pm1.2$ over the second highest preference in NLA-$z$).


\subsection{Intrinsic alignment parameters}
\begin{figure}
    \centering
    \includegraphics[width=\columnwidth]{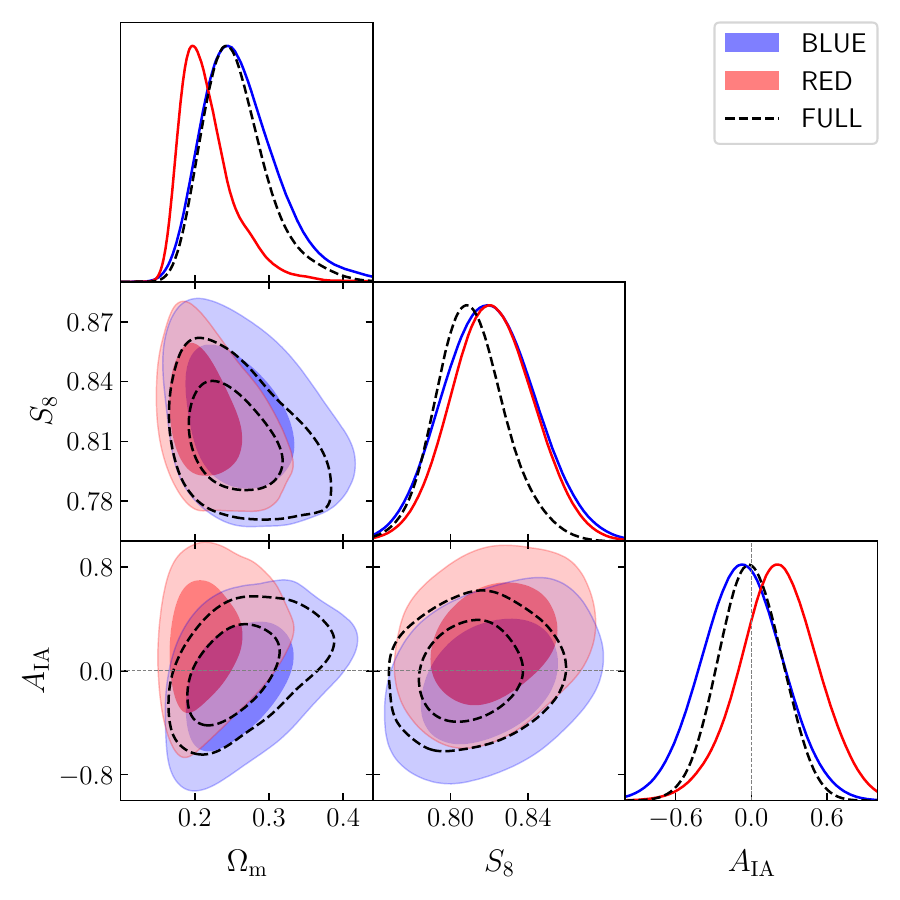}
    \caption{Intrinsic alignment amplitudes for the single parameter NLA model in the \textsc{red} (shaded), \textsc{blue} (shaded), and \textsc{full} (black, dashed) samples, with 1$\sigma$ and 2$\sigma$ contours plotted for each. We find the all samples to be consistent with no IA; $a_1=0$.}
    \label{fig:a1}
\end{figure}

\begin{figure}
    \centering
    \includegraphics[width=\columnwidth]{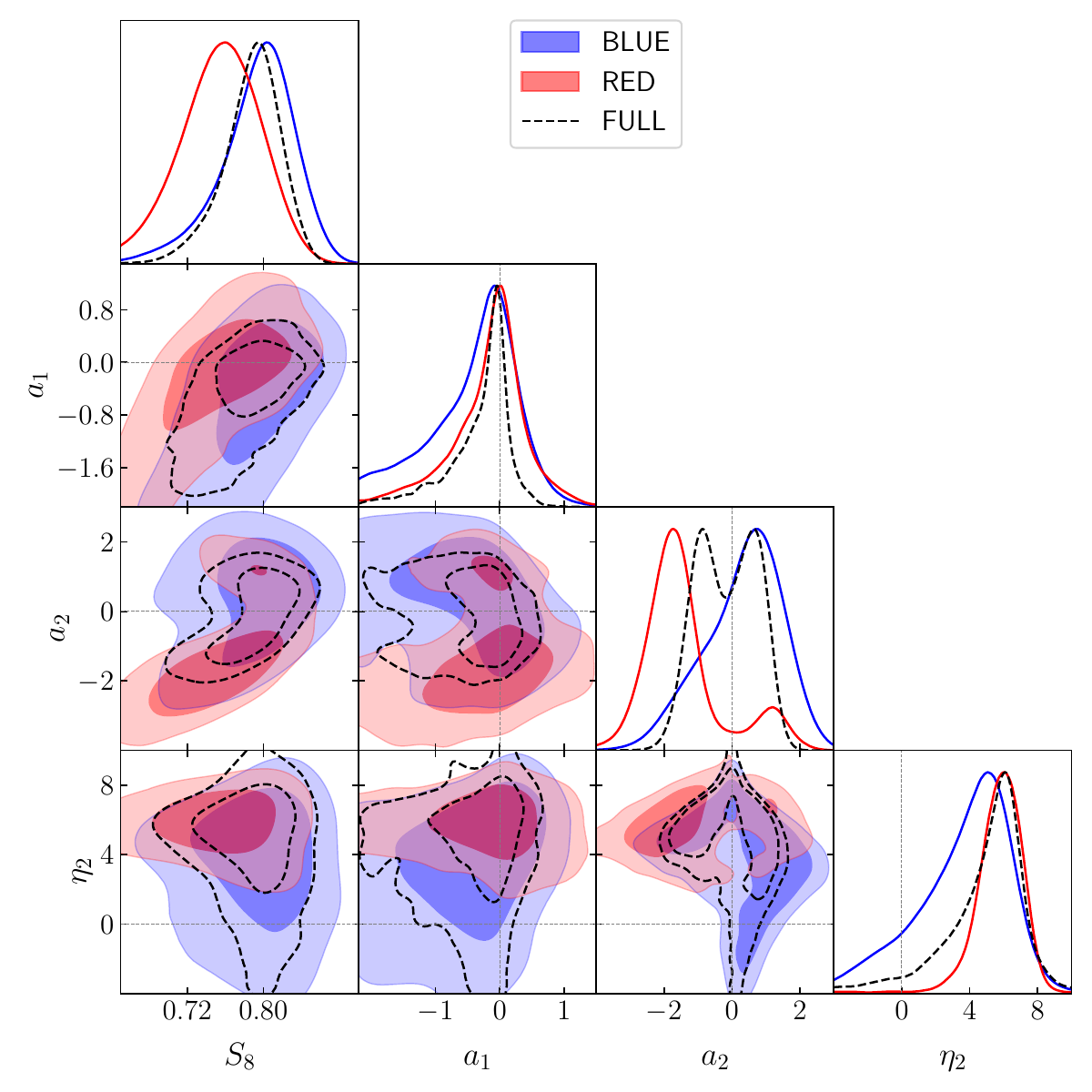}
    \caption{Constraints on the intrinsic alignment amplitudes, $a_{1, 2}$ and redshift evolution, $\eta_2$, TATT model parameters using the \textsc{red}, \textsc{blue}, and \textsc{full} (black) samples, with (1$\sigma$ and 2$\sigma$ contours). We find the \textsc{blue} sample is consistent with zero in both $a_1,a_2$, while the \textsc{red} sample prefers a negative, strongly redshift-evolving $a_2$ amplitude \rev{with negligible $a_1$}.}
    \label{fig:a1a2}
\end{figure}

The constraints on the alignment amplitude for NLA are shown in Fig.~\ref{fig:a1}. The \textsc{red} sample fits for a slightly positive $a_1$, though is consistent with zero at the level of 1-$\sigma$, and the blue and full samples both infer no intrinsic alignment.
The TATT model parameters are shown in Fig.~\ref{fig:a1a2} for all three samples. Again, the blue sample is consistent with zero IA; $a_1, a_2 = 0$ (dashed lines), \rev{where $a_2$ is the physically anticipated quadratic alignment mechanism for spirals}. In this more complex model, the \textsc{red} sample also finds zero alignment amplitude but prefers a negative tidal torquing amplitude $a_2$, with \rev{a very} strong redshift evolution \rev{that falls beyond qualitative expectation}. This $a_1 = 0$ deviates from the literature (e.g., \citealt{Samuroff_2019}), although we note that the impure \textsc{red} selection we make differs from past work. We find a degeneracy (1) in response to sign flips in $a_1 / a_2$ (as seen in \citealt{Amon_2022, Secco_2022}), as well as (2) between high alignment, low-$S_8$ and low alignment, high-$S_8$ models for the impure \textsc{red} sample specifically. No intrinsic alignment is only slightly preferred for this sample according to the Bayesian evidence ratios ($K = 1.3\pm1.2$ w.r.t. TATT), while it is substantially preferred in the other samples (see Table \ref{tab:cosmologyresults}).


\section{Conclusions}\label{sec:discussion}
The modeling of intrinsic alignments in weak lensing studies is challenging: the model choice between NLA and TATT tends to shift the cosmological constraints by >0.5$\sigma$ \citep[e.g.,][]{des+kids,Secco_2022, Bigwood, Arico2023}. Application of these models to cosmic shear is challenging as they have thus far only been tested against direct measurements on larger scales and lower redshifts than cosmic shear measurements are sensitive to (\rev{6 Mpc$/h$ and $k$$\sim$1 for NLA and 2 Mpc$/h$ and $k$$\sim$3  for TATT}).
As progress is made to build flexible baryon feedback models that allow for the analysis of the full scale dependence of shear measurements, the uncertainty in small scale IA modeling is more apparent. Furthermore, the more accurate TATT model incurs additional model parameters that weaken cosmological precision. Newer models \citep{Fortuna_2020,Vlah2020, Bakx2023, ChenKokron, Maion2024} could exacerbate that cost. 

Red and blue galaxy populations exhibit clear differences in their IA properties. Motivated by the fact that IA have not yet been detected for blue galaxies \citep[e.g.,][]{Johnston_2019}, which dominate a weak lensing sample, we explore a novel approach for IA mitigation: we select a \textsc{blue} `IA clean' sample, which we assume is negligibly impacted by IA based on current direct IA measurements. First, we assess the stability of cosmological constraints with different IA model choices when using the \textsc{blue} sample. Next, we compare the cosmological constraints and goodness of fit of three analyses: (1) the proposed approach to use all angular scales and a flexible feedback model, but limiting to the \textsc{blue} galaxies to mitigate IA, (2) using all angular scales of the DES Y3 \textsc{full} sample measurement and a flexible feedback model, with the TATT model for IA, and (3) the DES Y3 approach to eliminate the small-scale shear measurements of the \textsc{full} sample and instead, use the TATT model. 

In summary, we extract and analyse a high purity ($\gtrsim 97$\%) star-forming selection (\textsc{blue}) and the complementary color selected sample containing a large fraction ($\sim30$\%) of passive, quiescent galaxies (\textsc{red}). To do this, we repeat the DES Y3 data calibration and cosmic shear analysis. 
We find that:
\begin{itemize}
\setlength\itemsep{0.3em}

    \item  Restricting the analysis to the \textsc{blue} sample of star-forming galaxies produces cosmological constraints that are more stable to IA model choice (even with use of all angular scales), varying by $\sim$0.5$\sigma$ in $S_8$. In contrast, the impure \textsc{red} sample has a best fit cosmology that changes with choice of IA model by $\sim$1.5$\sigma$ in $S_8$. 

\item  The $\chi^2_{\rm min}$ are substantially improved for the \textsc{blue} sample (lower than that for \textsc{full}, \textsc{red}  by $\sim20,40$). Specifically, the proposed \textsc{blue} analysis of (1), that uses all angular scales of the measurement and \rev{does not have free} IA parameters gives a better fit to the data compared to the \textsc{full} sample analyzed with TATT, either with small scales (2) or without (3).

\item The constraints from the \textsc{blue} analysis are in better agreement with results from \textit{Planck} CMB in both $S_8$ and $\Omega_{\rm m}$ compared to the \textsc{full} sample analyses (2) and (3). We find ${S_8} = 0.822^{+0.019}_{-0.020}$ and $\Omega_{\rm m} = 0.268^{+0.031}_{-0.056}$ as our 68\% confidence regions, corresponding to 0.23 in $\delta S_8$.

\item Our proposed \textsc{blue} IA-mitigation strategy improves the cosmological precision compared to the DES Y3 analysis using the \textsc{full} sample and TATT model by a factor of $1.5$. That is, the approach of (1) to safely analyse the full angular scale extent with a flexible feedback model and a \textsc{blue} sample, where the latter circumvents the concern of IA model suitability, results in a reduced $S_8$ 68\% quantile compared to the \text{full} sample with TATT, using a feedback model (2) or scale cuts (3). 

\item The \textsc{blue} sample IA parameter constraints are consistent with zero amplitude in both NLA and TATT, and the analysis that neglects to model IA is strongly preferred. Consistent with \cite{Secco_2022}, we find this preference for all three samples, although it is very mild for the impure \textsc{red} sample over TATT (evidence ratio $\approx1.3$). When we do model alignment in the \textsc{red} sample, TATT finds strong tidal torquing (a negative $a_2$) with \rev{an unexpected,} extreme redshift evolution (high $\eta_2$).

\item All alignment models for the \textsc{red} sample prefer a low $\Omega_{\rm m}$ and higher $\chi^2$ than the uncontaminated \textsc{blue} sample, indicating that no model well describes the alignment in this color-redshift space, and that use of the full sample \rev{at affected scales} -- which contains these poorly modeled contaminants -- is ill-advised.
\end{itemize}

The next generation of weak lensing experiments \rev{has} arrived. The Vera Rubin \citep{lsst_science_overview}, Euclid \citep{euclid_overview}, and Roman \citep{roman_wfirst} Observatories will provide unprecedented quantities of data at unprecedented depth. Current weak lensing efforts have unveiled a deep need to understand and model systematics, among them IA as one of the chief confounding factors. A \textsc{blue} cosmic shear analysis, as described here, is a reliable way to account for our limited knowledge on IA. There are promising avenues to progress intrinsic alignment modeling -- for example, via direct measurements with spectroscopic surveys and shape measurement (e.g., \citealt{eboss}) or more flexible models that reduce constraining power (e.g., \citealt{Chen_2024}). However, given our present limited understanding and the ambiguity in model choice, we suggest that selection on blue galaxies allows us to perform precision weak lensing analyses with our surveys as they exist today \rev{while the ambiguity of IA model choice clarifies with observation}. This blue shear analysis is the simplest implementation of informative, observation driven priors on alignment that take into account galaxy properties like color and redshift, and lays the groundwork for more complex treatment in the future.

\section*{Data availability}
The public data release for DES Y3 is available at DESDM\footnote{\url{https://des.ncsa.illinois.edu/releases/y3a2}}. The blue selection of galaxies, data vector, and fiducial chain will be available for download at \url{https://jamiemccullough.github.io/data/blueshear/}.

\section*{Acknowledgements}
J. McCullough, A. Amon, E. Legnani, and D. Gruen completed the analysis and wrote the manuscript for this paper with the constructive feedback of A. Roodman. The other core authors, including O. Friedrich, N. MacCrann \& M. Becker, and J. Myles, contributed code and expertise to compute the covariance, to generate the DES Y3 image simulations and compute the shear calibration, and to calibrate the photometric redshifts, respectively. The document has been through internal review within the DES collaboration with S. Dodelson and S. Samuroff acting as internal reviewers with valuable expertise, as well as notable feedback throughout the process from J. Blazek and J. Prat. The remaining authors have made contributions to the Y3 Key Project analysis pipeline, including but not limited to DES instruments, data collection, processing and calibration, and various analysis pipelines.

Funding for the DES Projects has been provided by the DOE and NSF(USA), MEC/MICINN/MINECO(Spain), STFC(UK), HEFCE(UK). NCSA(UIUC), KICP(U. Chicago), CCAPP(Ohio State), 
MIFPA(Texas A\&M), CNPQ, FAPERJ, FINEP (Brazil), DFG(Germany) and the Collaborating Institutions in the Dark Energy Survey.

The Collaborating Institutions are Argonne Lab, UC Santa Cruz, University of Cambridge, CIEMAT-Madrid, University of Chicago, University College London, 
DES-Brazil Consortium, University of Edinburgh, ETH Z{\"u}rich, Fermilab, University of Illinois, ICE (IEEC-CSIC), IFAE Barcelona, Lawrence Berkeley Lab, 
LMU M{\"u}nchen and the associated Excellence Cluster Universe, University of Michigan, NFS NOIRLab, University of Nottingham, Ohio State University, University of 
Pennsylvania, University of Portsmouth, SLAC National Lab, Stanford University, University of Sussex, Texas A\&M University, and the OzDES Membership Consortium.

Based in part on observations at NSF Cerro Tololo Inter-American Observatory at NSF NOIRLab (NOIRLab Prop. ID 2012B-0001; PI: J. Frieman), which is managed by the Association of Universities for Research in Astronomy (AURA) under a cooperative agreement with the National Science Foundation.

The DES Data Management System is supported by the NSF under Grant Numbers AST-1138766 and AST-1536171. 
The DES participants from Spanish institutions are partially supported by MICINN under grants PID2021-123012, PID2021-128989 PID2022-141079, SEV-2016-0588, CEX2020-001058-M and CEX2020-001007-S, some of which include ERDF funds from the European Union. IFAE is partially funded by the CERCA program of the Generalitat de Catalunya.

We  acknowledge support from the Brazilian Instituto Nacional de Ci\^encia
e Tecnologia (INCT) do e-Universo (CNPq grant 465376/2014-2).

This manuscript has been authored by Fermi Research Alliance, LLC under Contract No. DE-AC02-07CH11359 with the U.S. Department of Energy, Office of Science, Office of High Energy Physics. 

This work has benefitted from support by the Deutsche Forschungsgemeinschaft (DFG, German Research Foundation) under Germany's Excellence Strategy – EXC-2094 – 390783311 and by the Bavaria California Technology Center (BaCaTec).

\section*{Affiliations}
$^{1}$ Department of Astrophysical Sciences, Princeton University, Peyton Hall, Princeton, NJ 08544, USA\\
$^{2}$ Kavli Institute for Particle Astrophysics \& Cosmology, P. O. Box 2450, Stanford University, Stanford, CA 94305, USA\\
$^{3}$ SLAC National Accelerator Laboratory, Menlo Park, CA 94025, USA\\
$^{4}$ Institut de F\'isica d’Altes Energies (IFAE), The Barcelona Institute of Science and Technology, Campus UAB, 08193 Bellaterra Barcelona, Spain\\
$^{5}$ Excellence Cluster ORIGINS, Boltzmannstr. 2, 85748 Garching, Germany\\
$^{6}$ University Observatory, Faculty of Physics, Ludwig-Maximilians-Universit\"at, Scheinerstr. 1, 81679 Munich, Germany\\
$^{7}$ University Observatory, Faculty of Physics, Ludwig-Maximilians-Universit\aa t, Scheinerstr. 1, 81679 Munich, Germany\\
$^{8}$ Department of Applied Mathematics and Theoretical Physics, University of Cambridge, Cambridge CB3 0WA, UK\\
$^{9}$ Argonne National Laboratory, 9700 South Cass Avenue, Lemont, IL 60439, USA\\
$^{10}$ Department of Physics, Carnegie Mellon University, Pittsburgh, Pennsylvania 15312, USA\\
$^{11}$ NSF AI Planning Institute for Physics of the Future, Carnegie Mellon University, Pittsburgh, PA 15213, USA\\
$^{12}$ Department of Physics, Northeastern University, Boston, MA 02115, USA\\
$^{13}$ Institut de F\'{\i}sica d'Altes Energies (IFAE), The Barcelona Institute of Science and Technology, Campus UAB, 08193 Bellaterra (Barcelona) Spain\\
$^{14}$ Department of Astronomy and Astrophysics, University of Chicago, Chicago, IL 60637, USA\\
$^{15}$ Nordita, KTH Royal Institute of Technology and Stockholm University, Hannes Alfv\'ens v\"ag 12, SE-10691 Stockholm, Sweden\\
$^{16}$ Center for Cosmology and Astro-Particle Physics, The Ohio State University, Columbus, OH 43210, USA\\
$^{17}$ Department of Physics, The Ohio State University, Columbus, OH 43210, USA\\
$^{18}$ Laborat\'orio Interinstitucional de e-Astronomia - LIneA, Rua Gal. Jos\'e Cristino 77, Rio de Janeiro, RJ - 20921-400, Brazil\\
$^{19}$ Observat\'orio Nacional, Rua Gal. Jos\'e Cristino 77, Rio de Janeiro, RJ - 20921-400, Brazil\\
$^{20}$ Institute of Space Sciences (ICE, CSIC),  Campus UAB, Carrer de Can Magrans, s/n,  08193 Barcelona, Spain\\
$^{21}$ Fermi National Accelerator Laboratory, P. O. Box 500, Batavia, IL 60510, USA\\
$^{22}$ Kavli Institute for Cosmological Physics, University of Chicago, Chicago, IL 60637, USA\\
$^{23}$ NASA Goddard Space Flight Center, 8800 Greenbelt Rd, Greenbelt, MD 20771, USA\\
$^{24}$ Instituto de F\'isica Gleb Wataghin, Universidade Estadual de Campinas, 13083-859, Campinas, SP, Brazil\\
$^{25}$ Centro de Investigaciones Energ\'eticas, Medioambientales y Tecnol\'ogicas (CIEMAT), Madrid, Spain\\
$^{26}$ Ruhr University Bochum, Faculty of Physics and Astronomy, Astronomical Institute, German Centre for Cosmological Lensing, 44780 Bochum, Germany\\
$^{27}$ Departments of Statistics and Data Sciences, University of Texas at Austin, Austin, TX 78757, USA\\
$^{28}$ NSF-Simons AI Institute for Cosmic Origins, University of Texas at Austin, Austin, TX 78757, USA\\
$^{29}$ Instituto de Astrofisica de Canarias, E-38205 La Laguna, Tenerife, Spain\\
$^{30}$ Universidad de La Laguna, Dpto. AstrofÃ­sica, E-38206 La Laguna, Tenerife, Spain\\
$^{31}$ Department of Physics and Astronomy, University of Pennsylvania, Philadelphia, PA 19104, USA\\
$^{32}$ Universit\'e Grenoble Alpes, CNRS, LPSC-IN2P3, 38000 Grenoble, France\\
$^{33}$ Department of Physics \& Astronomy, University College London, Gower Street, London, WC1E 6BT, UK\\
$^{34}$ Center for Astrophysics $\vert$ Harvard \& Smithsonian, 60 Garden Street, Cambridge, MA 02138, USA\\
$^{35}$ Santa Cruz Institute for Particle Physics, Santa Cruz, CA 95064, USA\\
$^{36}$ Department of Physics, University of Michigan, Ann Arbor, MI 48109, USA\\
$^{37}$ Institut d'Estudis Espacials de Catalunya (IEEC), 08034 Barcelona, Spain\\
$^{38}$ Institute of Cosmology and Gravitation, University of Portsmouth, Portsmouth, PO1 3FX, UK\\
$^{39}$ Jet Propulsion Laboratory, California Institute of Technology, 4800 Oak Grove Dr., Pasadena, CA 91109, USA\\
$^{40}$ Computer Science and Mathematics Division, Oak Ridge National Laboratory, Oak Ridge, TN 37831\\
$^{41}$ Brookhaven National Laboratory, Bldg 510, Upton, NY 11973, USA\\
$^{42}$ Department of Physics, ETH Zurich, Wolfgang-Pauli-Strasse 16, CH-8093 Zurich, Switzerland\\
$^{43}$ Department of Astronomy/Steward Observatory, University of Arizona, 933 North Cherry Avenue, Tucson, AZ 85721-0065, USA\\
$^{44}$ Department of Physics, University of Arizona, Tucson, AZ 85721, USA\\
$^{45}$ School of Physics and Astronomy, Cardiff University, CF24 3AA, UK\\
$^{46}$ Institut de Recherche en Astrophysique et Plan\'etologie (IRAP), Universit\'e de Toulouse, CNRS, UPS, CNES, 14 Av. Edouard Belin, 31400 Toulouse, France\\
$^{47}$ Institute of Theoretical Astrophysics, University of Oslo. P.O. Box 1029 Blindern, NO-0315 Oslo, Norway\\
$^{48}$ Department of Physics and Astronomy, University of Waterloo, 200 University Ave W, Waterloo, ON N2L 3G1, Canada\\
$^{49}$ George P. and Cynthia Woods Mitchell Institute for Fundamental Physics and Astronomy, and Department of Physics and Astronomy, Texas A\&M University, College Station, TX 77843,  USA\\
$^{50}$ Perimeter Institute for Theoretical Physics, 31 Caroline St. North, Waterloo, ON N2L 2Y5, Canada\\
$^{51}$ Max Planck Institute for Extraterrestrial Physics, Giessenbachstrasse, 85748 Garching, Germany\\
$^{52}$ Universit\"ats-Sternwarte, Fakult\"at f\"ur Physik, Ludwig-Maximilians Universit\"at M\"unchen, Scheinerstr. 1, 81679 M\"unchen, Germany\\
$^{53}$ Institute for Astronomy, University of Edinburgh, Edinburgh EH9 3HJ, UK\\
$^{54}$ Lawrence Berkeley National Laboratory, 1 Cyclotron Road, Berkeley, CA 94720, USA\\
$^{55}$ Jodrell Bank Center for Astrophysics, School of Physics and Astronomy, University of Manchester, Oxford Road, Manchester, M13 9PL, UK\\
$^{56}$ Instituto de Fisica Teorica UAM/CSIC, Universidad Autonoma de Madrid, 28049 Madrid, Spain\\
$^{57}$ LPSC Grenoble - 53, Avenue des Martyrs 38026 Grenoble, France\\
$^{58}$ Department of Physics and Astronomy, Pevensey Building, University of Sussex, Brighton, BN1 9QH, UK\\
$^{59}$ Physics Department, 2320 Chamberlin Hall, University of Wisconsin-Madison, 1150 University Avenue Madison, WI  53706-1390\\
$^{60}$ Australian Astronomical Optics, Macquarie University, North Ryde, NSW 2113, Australia\\
$^{61}$ Lowell Observatory, 1400 Mars Hill Rd, Flagstaff, AZ 86001, USA\\
$^{62}$ Department of Physics, University of Genova and INFN, Via Dodecaneso 33, 16146, Genova, Italy\\
$^{63}$ Department of Physics, Duke University Durham, NC 27708, USA\\
$^{64}$ Physics Department, Lancaster University, Lancaster, LA1 4YB, UK\\
$^{65}$ Center for Astrophysical Surveys, National Center for Supercomputing Applications, 1205 West Clark St., Urbana, IL 61801, USA\\
$^{66}$ Department of Astronomy, University of Illinois at Urbana-Champaign, 1002 W. Green Street, Urbana, IL 61801, USA\\
$^{67}$ Department of Astronomy, University of California, Berkeley,  501 Campbell Hall, Berkeley, CA 94720, USA\\
$^{68}$ School of Physics and Astronomy, University of Southampton,  Southampton, SO17 1BJ, UK\\
$^{69}$ Instituci\'o Catalana de Recerca i Estudis Avan\c{c}ats, E-08010 Barcelona, Spain\\
$^{70}$ Physics Department, William Jewell College, Liberty, MO, 64068\\
$^{71}$ School of Mathematics and Physics, University of Queensland,  Brisbane, QLD 4072, Australia\\
$^{72}$ Department of Physics, IIT Hyderabad, Kandi, Telangana 502285, India\\
$^{73}$ California Institute of Technology, 1200 East California Blvd, MC 249-17, Pasadena, CA 91125, USA\\
$^{74}$ Department of Physics and Astronomy, Stony Brook University, Stony Brook, NY 11794, USA\\
$^{75}$ University of Nottingham, School of Physics and Astronomy, Nottingham NG7 2RD, UK\\
$^{76}$ Departamento de F\'isica Matem\'atica, Instituto de F\'isica, Universidade de S\~ao Paulo, CP 66318, S\~ao Paulo, SP, 05314-970, Brazil\\
$^{77}$ Department of Physics, University of Oxford, Denys Wilkinson Building, Keble Road, Oxford OX1 3RH, UK\\
$^{78}$ Hamburger Sternwarte, Universit\"{a}t Hamburg, Gojenbergsweg 112, 21029 Hamburg, Germany\\
$^{79}$ Cerro Tololo Inter-American Observatory, NSF's National Optical-Infrared Astronomy Research Laboratory, Casilla 603, La Serena, Chile
\bibliographystyle{mnras} 
\bibliography{blueshear_bibliography}
\appendix
\section{Selection and Purity of \textsc{blue} Sample}
\label{app:color}

\begin{figure*}
    \centering
    \includegraphics{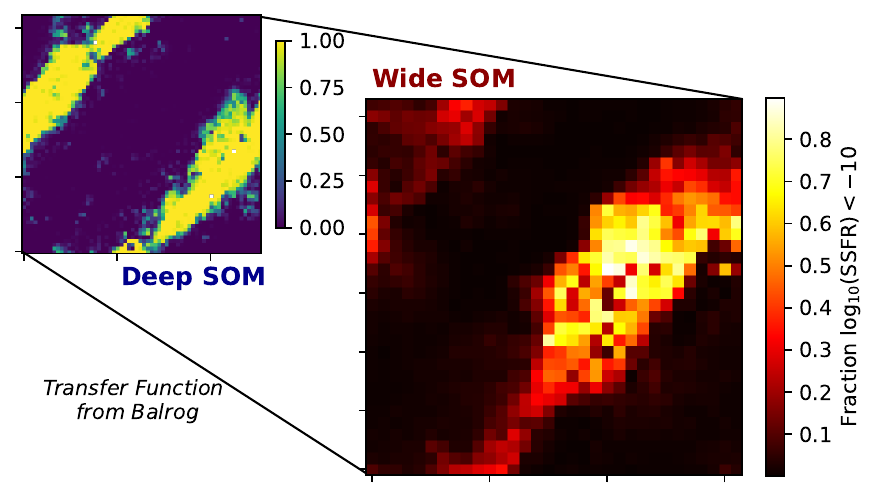}
    \caption{Depiction of the passive galaxy contamination of our selection, propagated from a cut on SSFR in the deep fields through the SOMPZ transfer function generated with an injected \textsc{balrog} sample. The contamination in our bin selections are described in Table \ref{tab:data_properties}.}
    \label{fig:red_purity}
\end{figure*}

As seen in Fig.~\ref{fig:red_purity}, we can use template fits run in the deep field from e.g., \textsc{bagpipes} and the \textsc{balrog} transfer function (see \citealt{Everett_2022}) to describe how realizations of red galaxies populate the lower-dimensional, $riz$ color space where we must perform our selection. The process of performing this inference on fractional purity, or the probability of being red $p(r)$, for a wide SOM color cell $\hat{c}$, that comprises part of a tomographic bin, $\hat{c}\in b$, follows as a function of deep field SOM color cells $c$, and our overall selection in those deep fields, $\hat{s}$, as
\begin{equation}
    p(r|b,\hat{s}) = \sum_{\hat{c}\in b}\sum_c p(r|c,\hat{s})p(c|\hat{c},\hat{s})p(\hat{c}|\hat{s}),
\end{equation}
where we define
\begin{itemize}
    \item $p(\hat{c}|\hat{s})$ as the overall abundances of galaxies in a given wide cell given our selection and can be written in terms of cell occupancies,
    \item $p(c|\hat{c},\hat{s})$ as our transfer function from \textsc{balrog},
    \item $p(r|c,\hat{s})$ as the determination from deep field photometry on how likely a galaxy in a given deep color cell is to be passive, non-star-forming, or \textit{red}.
\end{itemize}
The latter term can be estimated several ways given high resolution deep field data, and we presented three such potential metrics in Table \ref{tab:data_properties}. Our criteria for red contamination in each case is defined below:
\begin{itemize}
    \item \textsc{BAGPIPES}: Bayesian Analysis of Galaxies for Physical Inference and Parameter EStimation (\textsc{bagpipes}\footnote{\url{https://bagpipes.readthedocs.io}}) is a Bayesian SED fitting code that models the emission from galaxies in a broad range of wavelengths, fits these models to observational data, and generates multivariate posterior distributions of various galaxy physical properties (e.g., SSFR, stellar mass). Its ability to recover realistic properties has been studied comprehensively in simulations \citep{Carnall_2018}. We define our passive contamination by simply taking the galaxies where the best fit specific star formation rate, \texttt{ssfr\_best}, is very low, i.e.,
    \begin{equation}
        \mathrm{log}_{10}(\mathrm{SSFR}) < -10.
    \end{equation} 
    See the deep to wide mapping of this selection in Fig.~\ref{fig:red_purity}.
    \item \textsc{EAZY}: As a SED fitting code for photo-\textit{z} estimation, \textsc{EAZY} \citep{eazy_2008} supplies coefficients for a suite of semianalytically modeled basis templates. We define our red contamination as galaxies where the sum of best fit coefficients $A_t$ for the passive templates $t \in P$, is dominant over all templates $t$. Our exact threshold was chosen to naturally bisect the visible bimodal distribution:
    \begin{equation}
        \sum_{t\in P}A_t \geq 0.77 \sum_t A_t.
    \end{equation}
    \item Balmer: As this inference framework is being done where we have high confidence spectroscopic and narrow band redshifts (i.e., a well defined color-redshift relation), we can identify passive galaxies by their anticipated large balmer break color. With the deep SOM median redshift, we identify where the Balmer break at $(1+z)\times4000\AA$. We then take the filters (in deep, from \textit{ugrizJHK}) on either side of the balmer break (e.g., $u-r$, $i-Y$) as our color $BB_c$, where passive galaxies are liable to have a steep slope and select on 
    \begin{equation}
        BB_c > 2.0.
    \end{equation} 
    This particular method identifies a larger fraction of "red" contaminants, but agrees with the prior two methods on which cells are most contaminated.
\end{itemize}

\section{Data calibration}\label{sec:data}

To perform a measurement on cosmic shear, we must (1) understand the redshift distribution of the source galaxies and (2) understand the statistical relationship between a galaxy's shape and the true shear. To tackle (1), in Sec.~\ref{app:redshift}, we implement a hierarchical Bayesian inference scheme that informs photometric redshifts with well-known spectroscopic redshifts in limited fields. In order to characterize (2), in Sec.~\ref{app:shear}, we rely on a set of image simulations that can inject true galaxy shapes from a high-resolution imaging survey and apply shears to those shapes, before re-measuring them as the original instrument would. Due to blending, this results in small shifts of our redshift distributions described in Sec.~\ref{app:shear_redshift}. Lastly, we estimate covariance in Sec.~\ref{app:cov}.

\subsection{Redshift Calibration}
\label{app:redshift}
We follow the \textsc{sompz} procedure outlined in  \cite{Buchs_2019}, as implemented in \cite{Myles_2021} with a each tomographic bin split into a \textsc{red} and \textsc{blue} sample by their $r-z$ color (SOM cell assignment, see Fig.~\ref{fig:fidcolor} inset panels). The redshift distributions for each original tomographic bin, as well as the splits on color are seen in Fig.~\ref{fig:fidcolor}, following the same procedure for redshift inference from the same sources as \cite{Myles_2021} (see the fiducial \textit{SPC, PC, SC} sample).

As demonstrated in \cite{Amon_2022}, combining redshift inference from \textsc{sompz} with clustering redshifts provided no substantial shifts in cosmology. Therefore we simplify our analysis with a single realization of the tomographic $N(z)$s from \textsc{sompz} (i.e., no use of many realizations in \textsc{hyperrank}), and no addition of clustering redshifts (\textsc{WZ}) \rev{or shear ratios (SR)}. \rev{As the pipeline was extensively validated in DES Y3, we match the uncertainty on mean redshift per bin developed in \cite{Myles_2021}, and shift the mean redshifts according to our new sample selection.}

\subsection{Shear calibration}
\label{app:shear}
We repeat in part the analysis in \cite{MacCrann_2021} for DES Y3 on the same suite of image simulations to provide corrections in the form of a multiplicative bias on the \textsc{metacalibration} response. The simulations include catalogs with an applied constant shear ($\pm g_{1,2}$) and an applied shear in each of four bins of true redshift, or \textit{z}-slice, with bounds [0.0, 0.4, 0.7, 1.0, 3.0]. The latter allows for an estimation of the effect of redshift-dependent shear bias, which is largely believed to be the result of contributions from blending. This enters our analysis as a modification of the redshift distribution $n_{\gamma}(z)$, which will contribute per bin a mean multiplicative factor, \textit{m}, and a shift in mean redshift, $\Delta \bar{z}$.

We make use of a simplification of the alternative tomographic binning technique (see \citealt{MacCrann_2021} Sec.~3.7) to marginalize over for the final reported multiplicative bias. The two choices of binning are:
\begin{itemize}
    \item \textbf{Fiducial (fid)} : applies the same SOM cell - bin mapping as in the data analysis.
    \item \textbf{Abundance-weighting (weight)} : applies the same mapping as with the fiducial, but accounts for variation in the colors and distributions of galaxy populations between the simulations and data. This binning re-weights contributions on a cell-by-cell basis according to $N_{\mathrm{data},c} / N_{\mathrm{sim},c}$ (so-dubbed \textit{w}-match in \citealt{MacCrann_2021}).
\end{itemize}

As we will apply an overly conservative shear bias in this analysis, additional tomographic binning systems will not be applied, nor will the near-neighbor re-weighting (done to approximate clustering in the simulation suite as was computed for the DES Y3 gold sample analysis). This simplified approach is feasible only because shear calibration is not the driving systematic in the DES Y3 cosmology analysis. In lieu of these computations, we will simply double our estimation for uncertainty in \textit{m} and provide a conservative high limit for our shear bias.

\subsection{$N_{eff}(z)$ from Shear Calibration}
\label{app:shear_redshift}
\begin{figure*}
    \centering
    \includegraphics[width=\textwidth]{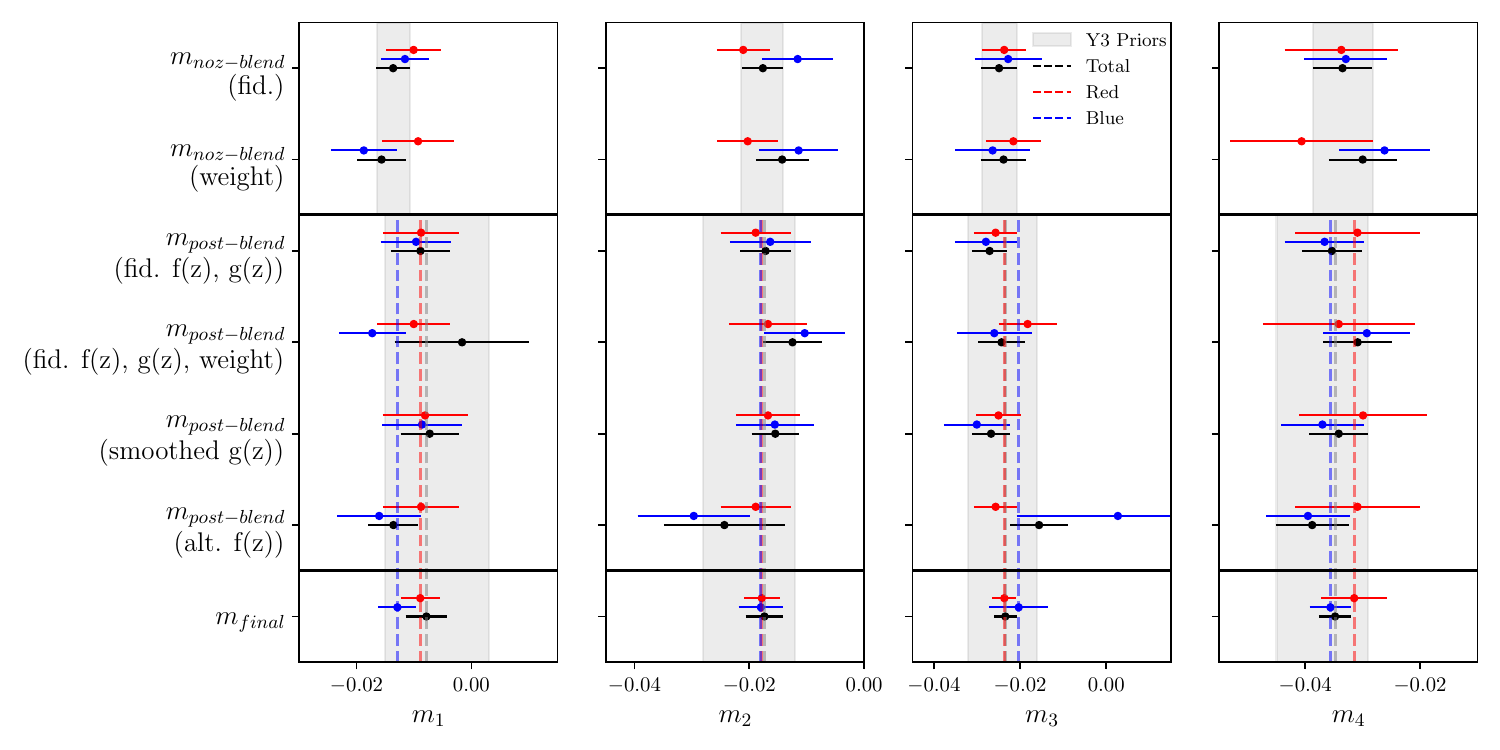}
    \caption{The recalculated shear bias measurements for \textsc{red} and \textsc{blue} galaxy splits following a simplified form of the analysis in \protect\cite{MacCrann_2021}. Our final shear priors on $m_i$ have the same width as the Y3 priors, but midpoints calculated as $m_{final}$.}
    \label{fig:msprior}
\end{figure*}

Following the shear calibration analysis methodology from DES Y3 in \cite{MacCrann_2021}, we produce samples of $f(z)$ and $g(z)$ from measurements in a suite of image simulations that modify our tomographic redshift distributions $N^i(z)$ in bin $i$ to adjust for blending contributions according to a prescription of
\begin{equation}
    N_{\gamma}^i(z) = \left(1 + f^i(z)\right)N^i(z) + g^i(z),
\end{equation}
where $N^i_{\gamma}$ is the new effective redshift distribution taking into account redshift dependent shear bias. We employed the fiducial (fid.) functional forms of $f$ and $g$ in \cite{MacCrann_2021} (i.e., their Sec.~5.5.1), alongside the alternative f(z) (alt., Student's-$t$) and the smoothed g(z). We display results on $m_i$ of applying these model fits to the simulations in Fig.~\ref{fig:msprior} and take the combined result for each tomographic bin as our new prior centers for $m_i$. The redshift distributions in Fig.~\ref{fig:fidcolor} contain these small modifications due to blending.


\setlength{\tabcolsep}{3pt}
\begin{table}
    \centering
    \begin{tabular}{c|c|c|c}
        \hline \hline
          & \textsc{full} &  \textsc{red} & \textsc{blue} \\
        \hline
        $m_{1,\mathrm{sim}}\times 100$ & $-1.362 \pm 0.296$ & $-1.006 \pm 0.473$ & $-1.156 \pm 0.419$ \\
        $m_{2,\mathrm{sim}}\times 100$ & $-1.764 \pm 0.359$ & $-2.106 \pm 0.460$ & $-1.158 \pm 0.623$ \\
        $m_{3,\mathrm{sim}}\times 100$ & $-2.488 \pm 0.417$ & $-2.369 \pm 0.512$ & $-2.277 \pm 0.777$ \\
        $m_{4,\mathrm{sim}}\times 100$ & $-3.350 \pm 0.513$ & $-3.371 \pm 0.989$ & $-3.292 \pm 0.722$ \\
        \hline
        $m_{1,\mathrm{final}}\times 100$ & $-0.783 \pm 0.361$ & $-0.890 \pm 0.339$ & $-1.288 \pm 0.329$ \\
        $m_{2,\mathrm{final}}\times 100$ & $-1.737 \pm 0.329$ & $-1.782 \pm 0.309$ & $-1.799 \pm 0.388$ \\
        $m_{3,\mathrm{final}}\times 100$ & $-2.343 \pm 0.262$ & $-2.366 \pm 0.277$ & $-2.031 \pm 0.680$ \\
        $m_{4,\mathrm{final}}\times 100$ & $-3.479 \pm 0.285$ & $-3.147 \pm 0.577$ & $-3.560 \pm 0.363$ \\
        \hline
    \end{tabular}
    \label{tab:appm}
    \caption{The calculated m priors in each bin from the image simulations (\textit{sim} or reported in \citealt{Amon_2022} as \textit{noz-blend}) and resulting final shear bias $m$, when the simulation calibration is applied to DES data.}
\end{table}

\subsection{Covariance}
\label{app:cov}
We follow our modeling of the analytic covariance matrix from the analysis in \cite{oliver_cov} for DES Y3. The calculations are carried out using CosmoCov \citep{cosmocov} for our split samples at the best-fit cosmology from the DES Y3 "3$\times$2pt" analysis that combined cosmic shear, galaxy-galaxy lensing, and clustering in \cite{3x2pt}. This covariance matrix was computed \rev{with tomographic effective number densities and average per-component shape noise quoted in Table~\ref{tab:data_properties} and calibrated redshift distributions} for all samples: \textsc{full, blue, red}.

\section{Sampling Cosmology}
\label{app:cosmology}
\subsection{Cosmic shear}
We can model our two-point correlation functions in shear by first beginning with the three dimensional non-linear matter power spectrum. This is written in terms of the two-dimensional convergence power spectrum from lensing, $C^{i,j}_{\kappa}$, which can be defined at a given angular wave number $\ell$ as a decomposition into E-modes and B-modes,

\begin{equation}
\centering
\label{eqn:2ptP}
 \xi_{\pm}(\theta) = \sum_{\ell} \frac{2\ell+1}{4\pi} G^\pm_\ell(\cos \theta) \left(C_{\kappa,\textrm{EE}}^{ij}(\ell) \pm C_{\kappa, \textrm{BB}}^{ij}(\ell)\right) \, .
\end{equation}

While weak lensing will not contribute to $C_{\kappa, \textrm{BB}}$, intrinsic alignment can. Spherical harmonics in \cite{stebbins_1996} provide $G^{\pm}_{\ell}$. For ease of computing, we implement the flat-sky and Limber approximations \citep{extended_limber}, the latter of which is accurate at high multipoles. With this, we can relate our convergence power spectrum from lensing to the overall three-dimensional matter power spectrum $P(k,z)$:
\begin{equation}\label{eq:c_kappa}
    C_{\kappa}^{ij}(\ell) = \int_0^{\chi_{\rm H}}  P\left(k=\frac{\ell+0.5}{\chi(z)}, z\right) \frac{W_i(\chi)W_j(\chi)}{\chi^2}  d\chi\,,
\end{equation}
where $\chi$ is the comoving angular diameter distance, $z$ is the redshift, and $\chi_{\rm H}$ is the distance to the horizon. The power spectrum is modulated by $W_i(\chi)$, which describes how efficiently it is picked up by lensing in a given tomographic bin. We write these kernels for lensing efficiency as

\begin{equation}\label{eq:kernel}
W_i(\chi)= \frac{3H_0^2\Omega_{\rm m}\chi}{2c^2a(\chi)}\int_{\chi}^{\chi_{\rm H}} \, n_i(\chi') \frac{\chi'-\chi}{\chi'}  d\chi'\,,
\end{equation}
where $n_i$ is the normalized effective number density of galaxies, $a$ is the scale factor, and both can be written as functions of comoving distance $\chi$. 

We therefore have an expression that can relate our observed measurements (shear two-point correlation) to simulated, cosmologically dependent models of the full nonlinear matter power spectrum.

\subsection{Modeling IA}
The direct quantities that intrinsic alignment models like NLA and TATT must predict are in the space of the two-point function, where our data vector lives. Galaxies can be intrinsically correlated with their neighbors, but also anti-correlated with background galaxies experiencing shear from the massive structure the aligned galaxies share. We can write the angular power spectrum between redshift bins $i,j$ in terms of these gravitational, \textit{G}, and intrinsic, \textit{I}, shears:
\begin{equation}
\langle \epsilon^{\mathrm{obs},i}\epsilon^{\mathrm{obs},j}  \rangle  =
\langle \gamma^i_{\rm G} \gamma^j_{\rm G}\rangle + \langle \gamma^i_{\rm G} \gamma^j_{\rm I}\rangle + \langle \gamma^i_{\rm I} \gamma^j_{\rm G}\rangle + \langle \gamma^i_{\rm I} \gamma^j_{\rm I}\rangle,
\label{eqn:xiia}
\end{equation}
where any terms involving $\epsilon$ at nonzero separation average to zero. 
\subsection{Calculation and Sampling}

When running cosmology chains we use the COSMOSIS framework \citep{Zuntz:2015,McEwen_2016,Lewis_2000,Howlett_2012} and follow the optimal sampler settings derived in \cite{lemos&weaverdyck2023}, to use\textsc{PolyChord} \citep{polychord,polychord2}. These settings were further confirmed in the joint DES-KiDS analysis \citep{des+kids}.

We use \textsc{CAMB} to calculate the linear component of the matter power spectrum \citep{camb}, and \textsc{HMCode2020} for the non-linear correction \citep{mead_2020}. For simplicity, we fix two massless neutrino species and a third with the lowest mass permitted in oscillation experiments ($m_\nu$ = 0.06eV, e.g., \citealt{Patrignani_2016}).

We model the baryonic feedback effects according to a wider, higher prior than in previous studies \rev{(see Table \ref{tab:cosmo_priors_app}), and with more large-scale impact than accounted for in} DES Y3 \citep{amon_2022_s8}, by taking into account more recent analyses that find the suppression of the nonlinear power spectrum at small scales requires stronger feedback \citep{boryana,Bigwood}. We make use of a parameter within \textsc{HMCode2020}, $\Theta_{\rm AGN}$, to model this contribution to the correlation function \citep{mead_2020}. As a result of this generous prior, and our intention to study competing models of intrinsic alignment, we examine all angular scales of the correlation function.

\begin{figure*}
    \centering
    \includegraphics[width=0.98\textwidth]{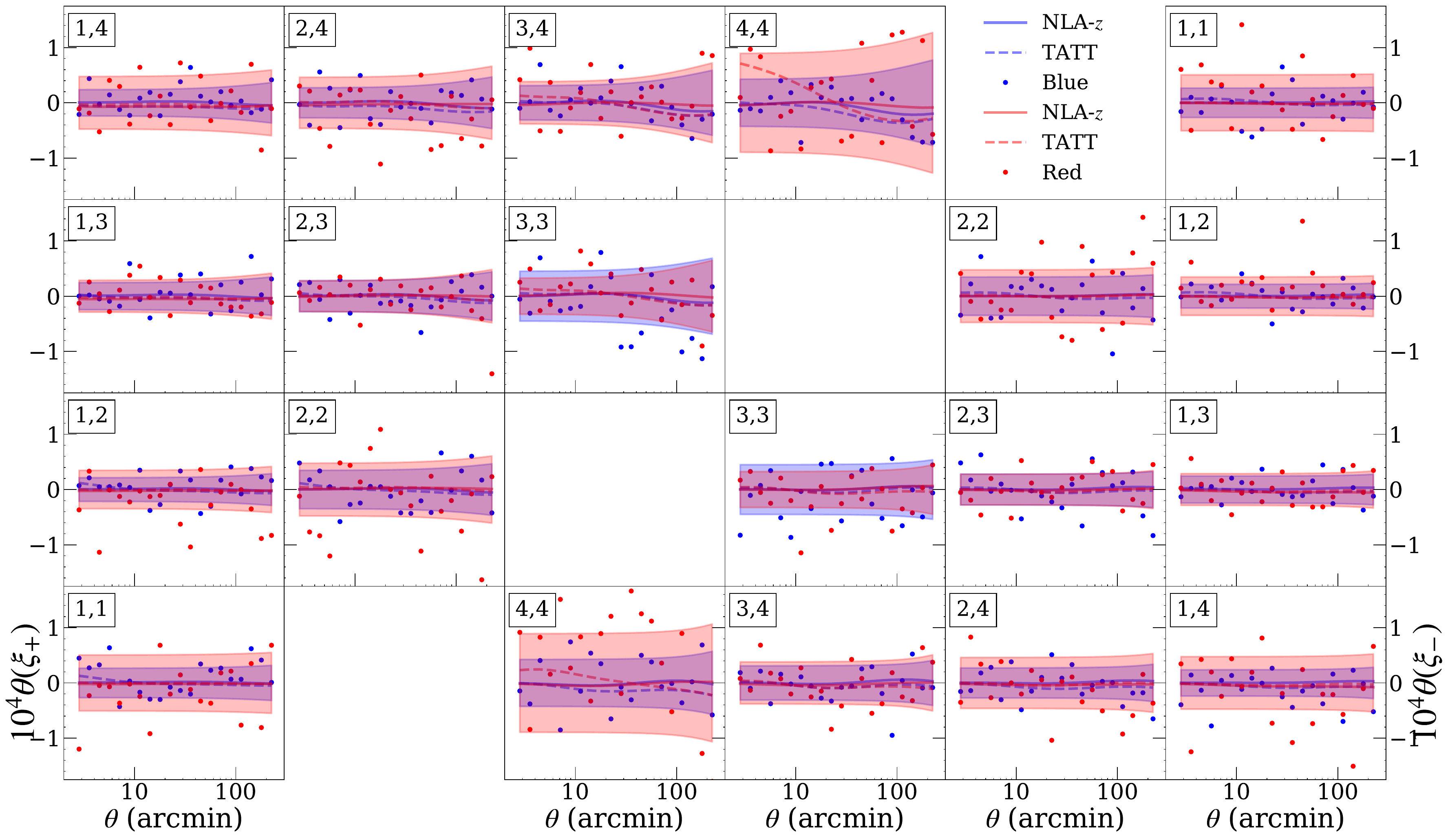}
    \caption{Cosmic shear two-point correlation residuals between the measurements for the \textsc{blue} and \textsc{red} samples compared to the `no IA' best fit model. The shaded envelope represents the square root of the diagonal of the analytic covariance matrix.  The solid and dashed lines represent residuals computed with the best fit using the NLA and TATT IA models. The differences between the best fits using different IA models are small and within the measurement uncertainty. The most notable shift compared to `no IA' is with the TATT model in the highest redshift bin.} 
    \label{fig:cosmicshear_resid}
\end{figure*}

\setlength{\tabcolsep}{1pt}
\begin{table*}
    \caption{Summary of cosmological, observational and astrophysical parameters and priors used in the analysis. In the case of flat priors, indicated by square bracketsm the prior bound to the range indicated in the `value' column. Gaussian priors are indicated by brackets, and are described as (mean, 1-$\sigma$ width).}   
    \label{tab:cosmo_priors_app}
\begin{tabular}{cccc}
\hline
\hline
Parameter & \multicolumn{2}{c}{Value} \tabularnewline
\hline 
\bf{Cosmological} \tabularnewline
$\Omega_{\rm m}$, Total matter density & \multicolumn{2}{c}{[0.1, 0.9]} \tabularnewline
 $\Omega_{\rm b}$, Baryon density &  \multicolumn{2}{c}{[0.03, 0.07]} \tabularnewline
$A_{\rm s}$, Scalar spectrum amplitude  &  \multicolumn{2}{c}{[0.5, 5.0] $\times 10^{-9}$}\tabularnewline
$h$, Hubble parameter  &  \multicolumn{2}{c}{[0.55, 0.91]}  \tabularnewline
$n_{\rm s}$, Spectral index  & \multicolumn{2}{c}{[0.87,1.07]} \tabularnewline
\hline 
\bf{Baryon Feedback} \tabularnewline
$\mathrm{log_{10}}(T_{\rm AGN})$, Strength of AGN feedback & \multicolumn{2}{c}{[7.6, 8.3]} \tabularnewline
\hline 
\bf{Intrinsic alignment} \tabularnewline
$a_1$, Tidal alignment amplitude (NLA) &  \multicolumn{2}{c}
{$[-5,5]$} \tabularnewline
$\eta_1$, Tidal alignment $z$-index     & \multicolumn{2}{c}{$[-5,5]$} \tabularnewline
$a_2$, Tidal torque amplitude      &  \multicolumn{2}{c}{$[-5,5]$} \tabularnewline
$\eta_2$, Tidal torque $z$-index      &  \multicolumn{2}{c}{$[-5,10]$} \tabularnewline
$b_{\rm ta}$, Tidal alignment bias  &  \multicolumn{2}{c}{$[0,2]$} \tabularnewline 
\hline

\bf{Observational} &\textbf{\textsc{blue}}&\textbf{\textsc{red}}\tabularnewline
$\Delta z^1$, Redshift calibration uncertainty 1 &  ( 0.0, 0.018 )& ( 0.0, 0.018 ) \tabularnewline
$\Delta z^2$, Redshift calibration uncertainty 2 &  ( 0.0, 0.015 )& ( 0.0, 0.015 ) \tabularnewline
$\Delta z^3$, Redshift calibration uncertainty 3 &  ( 0.0, 0.011 )& ( 0.0, 0.011 ) \tabularnewline
$\Delta z^4$, Redshift calibration uncertainty 4  &  ( 0.0, 0.017 )& ( 0.0, 0.017 ) \tabularnewline
$m^1$, Shear calibration uncertainty 1 &  ( $-0.013$, 0.009 )& ( $-0.009$, 0.009 )\tabularnewline
$m^2$, Shear calibration uncertainty 2 &  ( $-0.018$, 0.008 ) & ( $-0.018$, 0.008 )\tabularnewline
 $m^3$, Shear calibration uncertainty 3 &  ( $-0.020$, 0.008 )& ( $-0.024$, 0.008 )\tabularnewline
$m^4$, Shear calibration uncertainty 4 &  ( $-0.036$, 0.008 )& ( $-0.032$, 0.008 )\tabularnewline
\hline 
\end{tabular}
\end{table*}
\setlength{\mytabcolsep}{\tabcolsep} 
In Table \ref{tab:cosmo_priors_app}, we record priors on model the uncertainty in mean redshift and the shear calibration for each tomographic bin $i$ as the free parameters $\Delta z^i$ and $m^i$. We hit prior boundaries in TATT parameter $\eta_2$, and thus extend that prior from [-5,5] to [-5,10], deviating from the fiducial Y3 analysis in \cite{Amon_2022, Secco_2022}. While the physical implication of a redshift evolution term this high is unclear, we expect this is likely a degenerate feature within the TATT model as fitted to our data. We demonstrate residuals of our model fits with respect to the preferred no intrinsic alignment model in Fig.~\ref{fig:cosmicshear_resid}.

\begin{figure*}
    \includegraphics[width=0.68\textwidth]{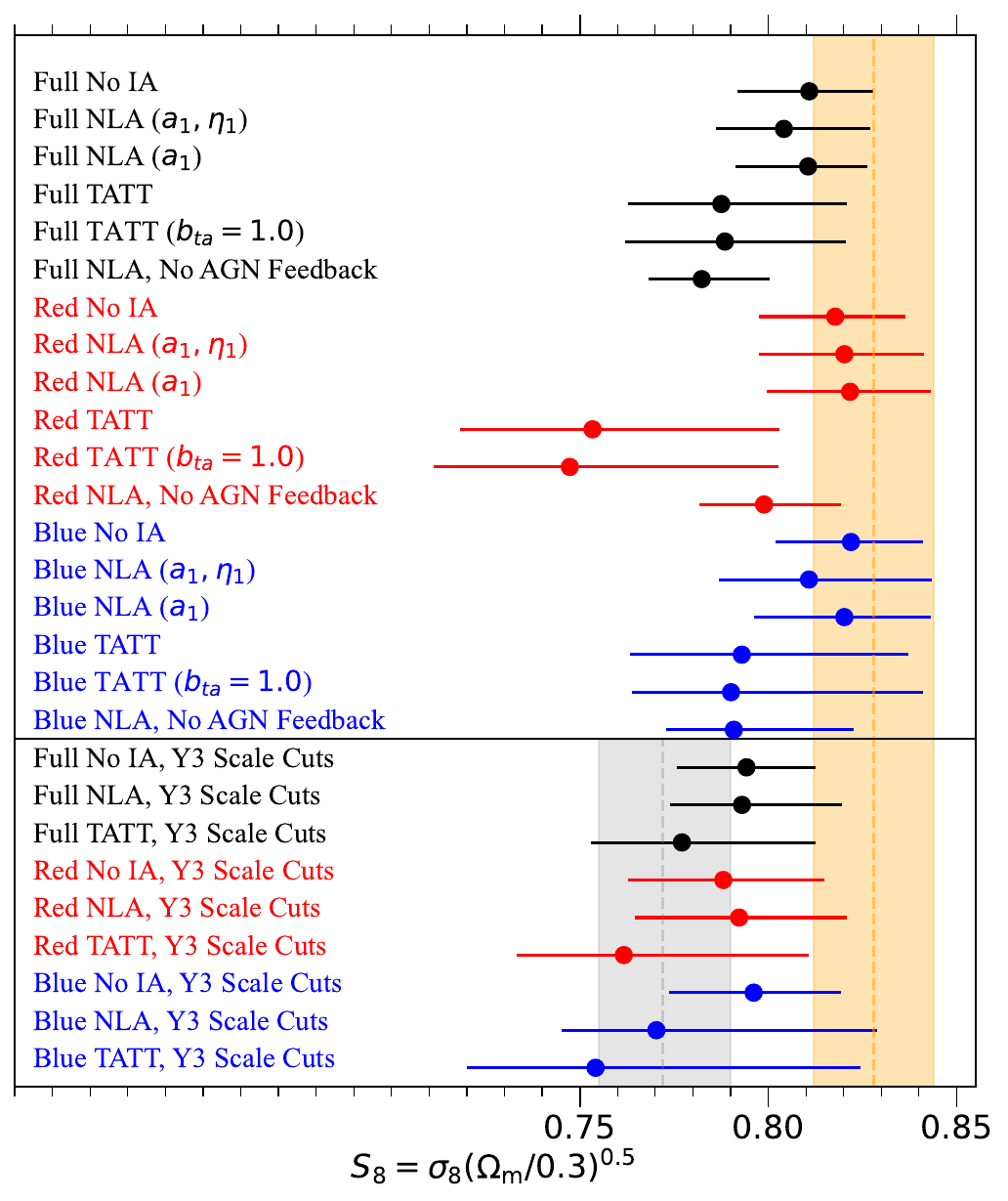}
    \hspace{3pt}
    \includegraphics[width=0.288\textwidth]{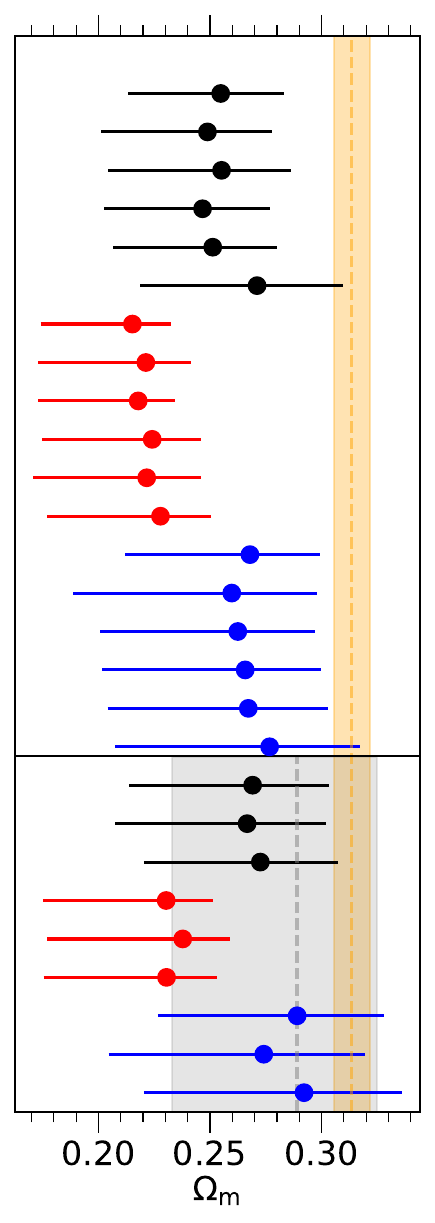}
    \caption{Summary statistics for each cosmological model fit, with the DES Y3 $\Lambda$CDM-Optimized result in shaded gray, and the \textit{Planck} results \citep{EfstathiouGratton:2021} in shaded orange. The 68\% confidence regions (1-$\sigma$) are reported for each model fit. Note the DES Y3 fiducial results use a slightly different tomographic binning and prior range.}
    \label{fig:datasummary}
\end{figure*}

Our full results are recorded in Table~\ref{tab:full_table} for all sampled models, and are summarized in Fig.~\ref{fig:datasummary} for $S_8$ and $\Omega_{\rm m}$.

\subsection{IA-photo-$z$ interplay}
\label{app:ia_pz}
Systematic uncertainty if not properly modeled and understood can be absorbed in other nuisance parameters. IA parameters can be degenerate with uncertain photometric redshift calibration. The effects of this can be visible in posterior distributions for shifts in the mean redshift of each tomographic bin, and intrinsic alignment parameter fits, as have been explored in \cite{Leonard_2024}. 

\begin{figure*}
    \centering
    \includegraphics[width=0.98\textwidth]{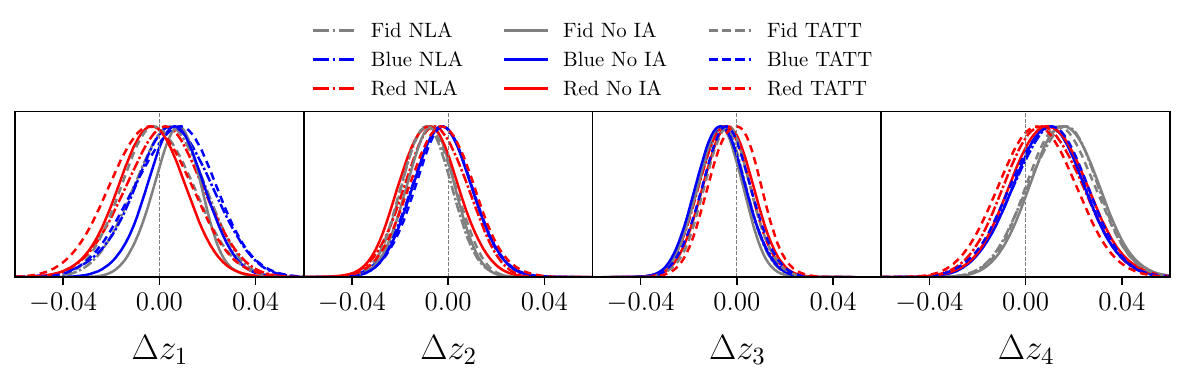}
    \caption{Posterior distributions for our major samples (\textsc{full}, impure \textsc{red}, pure \textsc{blue}) and alignment models (TATT, NLA, No IA) for the mean redshift per tomographic bin. We do not see any major shifts in $\bar{z}_i$ beyond 1-$\sigma$, indicating low degeneracy with IA.}
    \label{fig:dz_1d}
\end{figure*}

We do not find any major shifts in mean redshift, as depicted in Fig.~\ref{fig:dz_1d}, though the first redshift bin tends to prefer a slightly low $\Delta\bar{z}_1$ in the \textsc{red} sample and a slightly high $\Delta\bar{z}_1$ in the \textsc{blue}, these shifts are cohesive independent of IA model choice. We conclude that any interplay between IA modeling and photo-$z$ calibration is small.

\setlength{\tabcolsep}{6pt}
\begin{table*}
\resizebox{\textwidth}{!}{
\begin{tabular}{cccccccccccc}
\hline
\hline
\\ Model & $S_{8, mean}$ & $\Omega_{m, mean} $ & \makecell{Evid.\\Ratio} & $\chi_{\mathrm{min}}^2$  & $\chi_{\mathrm{red}}^2$ & $p$ & $\delta S_8$ & $\delta \Omega_m$ & $\sqrt{\delta S_8^2 + \delta \Omega_m^2}$\\
\hline
 \textsc{full} NLA-$z$ ($a_1$) & $ 0.811^{+0.016}_{-0.019} $ & $0.255^{+0.031}_{-0.051} $& {1.0} & {403.3} & {1.01} & {0.4093} & {0.720} & {1.292} & {1.479} \\ 
  \textsc{full} NLA ($a_1,\eta_1$) & $ 0.804^{+0.023}_{-0.018} $ & $0.249^{+0.029}_{-0.048} $& {1.5} & {404.0} & {1.02} & {0.3934} & {0.866} & {1.511} & {1.742} \\ 
  \textbf{(2) \textsc{full} TATT} & $ 0.788^{+0.033}_{-0.025} $ & $0.247^{+0.030}_{-0.044} $& {0.8} & {399.5} & {1.01} & {0.4271} & {1.159} & {1.722} & {2.076} \\ 
  \textsc{full} TATT ($b_{ta} = 1.0$) & $ 0.788^{+0.032}_{-0.027} $ & $0.251^{+0.029}_{-0.045} $& {0.9} & {399.4} & {1.01} & {0.4282} & {1.169} & {1.572} & {1.959} \\ 
  \textsc{full} No IA & $ 0.811^{+0.017}_{-0.019} $ & $0.255^{+0.028}_{-0.042} $& {20.4} & {403.9} & {1.01} & {0.4077} & {0.716} & {1.528} & {1.687} \\ 
  \textsc{full} NLA, No AGN Feedback & $ 0.782^{+0.018}_{-0.014} $ & $0.271^{+0.039}_{-0.052} $& {0.3} & {407.0} & {1.03} & {0.3535} & {1.931} & {0.887} & {2.125} \\ 
  \textsc{full} NLA, Y3 Scale Cuts\textsuperscript{\textdagger} & $ 0.793^{+0.027}_{-0.019} $ & $0.267^{+0.035}_{-0.059} $& {1.0} & {299.8} & {1.11} & {0.1029} & {1.208} & {0.917} & {1.517} \\ 
  \textsc{full} No IA, Y3 Scale Cuts\textsuperscript{\textdagger} & $ 0.794^{+0.018}_{-0.019} $ & $0.269^{+0.034}_{-0.055} $& {9.0} & {298.9} & {1.10} & {0.1172} & {1.374} & {0.965} & {1.679} \\ 
  \textbf{(3) \textsc{full} TATT, Y3 Scale Cuts}\textsuperscript{\textdagger} & $ 0.777^{+0.035}_{-0.024} $ & $0.273^{+0.035}_{-0.052} $& {0.9} & {293.1} & {1.09} & {0.1399} & {1.418} & {0.900} & {1.680} \\ 
 \hline 
  \textsc{red} NLA-$z$ ($a_1$) & $ 0.822^{+0.022}_{-0.022} $ & $0.218^{+0.017}_{-0.045} $& {1.0} & {426.4} & {1.07} & {0.1530} & {0.237} & {2.395} & {2.407} \\ 
  \textsc{red} NLA ($a_1,\eta_1$) & $ 0.820^{+0.021}_{-0.023} $ & $0.221^{+0.020}_{-0.048} $& {0.7} & {424.5} & {1.07} & {0.1638} & {0.277} & {2.237} & {2.255} \\ 
  \textsc{red} TATT & $ 0.753^{+0.050}_{-0.035} $ & $0.224^{+0.022}_{-0.049} $& {3.7} & {419.2} & {1.06} & {0.1926} & {1.574} & {2.163} & {2.675} \\ 
  \textsc{red} TATT ($b_{ta} = 1.0$) & $ 0.747^{+0.055}_{-0.036} $ & $0.222^{+0.024}_{-0.051} $& {2.9} & {418.6} & {1.06} & {0.1987} & {1.587} & {2.075} & {2.612} \\ 
  \textsc{red} No IA & $ 0.818^{+0.019}_{-0.020} $ & $0.215^{+0.017}_{-0.041} $& {5.9} & {425.1} & {1.07} & {0.1673} & {0.393} & {2.759} & {2.787} \\ 
  \textsc{red} NLA, No AGN Feedback & $ 0.799^{+0.020}_{-0.017} $ & $0.228^{+0.023}_{-0.051} $& {0.8} & {426.4} & {1.07} & {0.1487} & {1.169} & {1.977} & {2.297} \\ 
  \textsc{red} NLA, Y3 Scale Cuts\textsuperscript{\textdagger} & $ 0.792^{+0.029}_{-0.028} $ & $0.238^{+0.021}_{-0.061} $& {1.0} & {302.8} & {1.12} & {0.0830} & {1.071} & {1.433} & {1.789} \\ 
  \textsc{red} No IA, Y3 Scale Cuts\textsuperscript{\textdagger} & $ 0.788^{+0.027}_{-0.025} $ & $0.230^{+0.021}_{-0.055} $& {5.1} & {302.9} & {1.12} & {0.0891} & {1.344} & {1.788} & {2.236} \\ 
  \textsc{red} TATT, Y3 Scale Cuts\textsuperscript{\textdagger} & $ 0.762^{+0.049}_{-0.029} $ & $0.231^{+0.023}_{-0.055} $& {0.3} & {301.5} & {1.12} & {0.0781} & {1.414} & {1.728} & {2.232} \\ 
 \hline 
  \textsc{blue} NLA-$z$ ($a_1$) & $ 0.820^{+0.023}_{-0.024} $ & $0.262^{+0.035}_{-0.062} $& {1.0} & {380.2} & {0.96} & {0.7257} & {0.277} & {0.952} & {0.991} \\ 
  \textsc{blue} NLA ($a_1,\eta_1$) & $ 0.811^{+0.032}_{-0.024} $ & $0.260^{+0.038}_{-0.071} $& {1.6} & {380.6} & {0.96} & {0.7144} & {0.491} & {0.912} & {1.036} \\ 
  \textsc{blue} TATT & $ 0.793^{+0.044}_{-0.030} $ & $0.266^{+0.034}_{-0.064} $& {0.6} & {375.4} & {0.95} & {0.7540} & {0.788} & {0.872} & {1.175} \\ 
  \textsc{blue} TATT ($b_{ta} = 1.0$) & $ 0.790^{+0.051}_{-0.026} $ & $0.267^{+0.036}_{-0.063} $& {0.5} & {375.9} & {0.95} & {0.7478} & {0.798} & {0.850} & {1.166} \\ 
  \textbf{(1) \textsc{blue} No IA} & $ 0.822^{+0.019}_{-0.020} $ & $0.268^{+0.031}_{-0.056} $& {12.7} & {380.9} & {0.96} & {0.7225} & {0.232} & {0.951} & {0.978} \\ 
  \textsc{blue} NLA, No AGN Feedback & $ 0.791^{+0.032}_{-0.018} $ & $0.277^{+0.041}_{-0.069} $& {1.1} & {381.6} & {0.96} & {0.7022} & {1.094} & {0.611} & {1.253} \\ 
  \textsc{blue} NLA, Y3 Scale Cuts\textsuperscript{\textdagger} & $ 0.770^{+0.059}_{-0.025} $ & $0.274^{+0.045}_{-0.070} $& {1.0} & {259.0} & {0.96} & {0.6743} & {1.102} & {0.637} & {1.273} \\ 
  \textsc{blue} No IA, Y3 Scale Cuts\textsuperscript{\textdagger} & $ 0.796^{+0.023}_{-0.022} $ & $0.289^{+0.039}_{-0.062} $& {6.7} & {259.5} & {0.96} & {0.6822} & {1.128} & {0.450} & {1.215} \\ 
  \textsc{blue} TATT, Y3 Scale Cuts\textsuperscript{\textdagger} & $ 0.754^{+0.070}_{-0.034} $ & $0.292^{+0.044}_{-0.072} $& {1.1} & {252.9} & {0.94} & {0.7381} & {1.216} & {0.347} & {1.264} \\ 
 
\hline
\end{tabular}
}
\footnotetext[0]{\textsuperscript{\textdagger}Chains with optimized Y3 scale cuts (defined in \citealt{Amon_2022, Secco_2022}) do not model AGN feedback to compare to prior results, $N_{\rm data} = 273$. }
\caption{Statistics for the comprehensive suite of Markov Chain Monte Carlo (MCMC) chains run with variations in model choice and sample of galaxies for cosmic shear, a subset of this table is displayed in Table \ref{tab:cosmologyresults}, where we provide column definitions.}
\label{tab:full_table}
\end{table*}

\setlength{\mytabcolsep}{\tabcolsep} 

\end{document}